\documentclass[a4paper, 11pt, oneside]{article}
\pdfoutput=1
\usepackage{float}
\usepackage{jheppub}
\usepackage{epstopdf}

\newcommand{\nc}{\newcommand}

\newlength{\absize}
\nc{\non}{\nonumber}
\nc{\hc}{\hbox {H.c.}} 
\nc{\noi}{\noindent}
\nc{\barx}{\bar{x}}
\nc{\pbarn}{\;\hbox {pb}}
\nc{\fbarn}{\;\hbox {fb}}
\nc{\lsp}{\;\;\;\;\;}
\nc{\Lsp}{\;\;\;\;\;\;\;\;\;\;}  
\nc{\LLsp}{\lspace \lspace}
\nc{\lra}{\longrightarrow}
\nc{\beq}{\begin{equation}}  \nc{\eeq}{\end{equation}}
\nc{\bea}{\begin{eqnarray}}  \nc{\eea}{\end{eqnarray}}
\nc{\baa}{\begin{array}}     \nc{\eaa}{\end{array}}
\nc{\bit}{\begin{itemize}}   \nc{\eit}{\end{itemize}}
\nc{\ben}{\begin{enumerate}} \nc{\een}{\end{enumerate}}
\nc{\bce}{\begin{center}}    \nc{\ece}{\end{center}}
\nc{\bpm}{\begin{pmatrix}}   \nc{\epm}{\end{pmatrix}}
\nc{\bvt}{\begin{verbatim}}  \nc{\evt}{\end{verbatim}}
\def\lsim{\mathrel{\raise.3ex\hbox{$<$\kern-.75em\lower1ex\hbox{$\sim$}}}}
\def\gsim{\mathrel{\raise.3ex\hbox{$>$\kern-.75em\lower1ex\hbox{$\sim$}}}}

\def\mcal{{\cal M}}

\def\ocal{{\cal O}}

\def\zcal{{\cal Z}}

\def\gev{\;\hbox{GeV}}

\nc{\tanb}{\tan\beta}
\nc{\mch}{M_{H^\pm}}

\def\mch{M_{H^\pm}}

\nc{\for}{\lsp {\rm for} \lsp}
\nc{\andd}{\lsp {\rm and} \lsp}

\renewcommand{\Re}{\mbox{Re\thinspace}}
\renewcommand{\Im}{\mbox{Im\thinspace}}
\newcommand{\half}{{\textstyle\frac{1}{2}}}

\allowdisplaybreaks 

\def\i11{{\mathbbm 1}}

\title{Diagnosing CP properties of the 2HDM}
\author[a]{B. Grzadkowski,}
\affiliation[a]{Faculty of Physics, University of Warsaw, Ho\.za 69, 00-681 Warsaw, Poland}
\author[b]{O. M. Ogreid,}
\affiliation[b]{Bergen University College, Postboks 7030, N-5020 Bergen, Norway}
\author[c]{P. Osland}
\affiliation[c]{Department of Physics,
University of Bergen, Postboks 7803, N-5020 Bergen, Norway}
\emailAdd{bohdan.grzadkowski@fuw.edu.pl}
\emailAdd{omo@hib.no}
\emailAdd{Per.Osland@ift.uib.no}
\date{\today}

\abstract{We have investigated a Two-Higgs-Doublet Model (2HDM), 
focusing on CP violation. Various scenarios with spontaneous and explicit breaking 
of CP have been considered. Some features of CP violation related to a choice of the 
basis for the two Higgs doublets have been discussed and clarified. Regions in the physical
parameter space corresponding to spontaneous and explicit CP violation have been located
and discussed. The possibility to determine parameters of the scalar potential with no reference to Yukawa couplings has been considered and an unavoidable ambiguity has been found. The issue of disentangling spontaneous and explicit
CP violation has been investigated.}


\begin{document}
\begin{flushright}
NORDITA-2013-79\
\end{flushright}

\maketitle

\flushbottom

\section{Introduction}

We investigate and discuss in detail the issue of CP violation (CPV)
in the scalar sector of the Two-Higgs-Doublet Model (2HDM). In spite of the existing rich 
literature (see, for example \cite{Branco:2011iw}) we believe that it is worth revisiting this issue with particular emphasis
on the possibility of spontaneous CP violation from a phenomenological point of view.
The ultimate goal of our study is to identify observables which will distinguish between explicit (ECPV) and spontaneous CP violation (SCPV) without reconstructing the full potential. 
For early literature on this question, see \cite{Lavoura:1994yu}.
The aim of the present paper is more modest: we will determine and display regions of explicit and spontaneous CP violation in the physical
parameter space of the model, i.e., in terms of parameters used directly in coupling
constants of mass eigenstates, such as mixing angles of neutral
scalars, masses, and vacuum expectation values (VEVs).

In general, the parameter regions where spontaneous CP violation occurs are embedded in regions of explicit CP violation, forming  lower-dimensional sub-spaces or manifolds. 
They can only be located where the potential has two minima of equal depth. However, the converse is not true: not all locations where there are two minima of equal depth correspond to spontaneous CP violation \cite{Ivanov:2007}. 
Thus, if the potential $V$ has two minima labeled $A$ and $B$, spontaneous CP violation may only occur at the manifolds constituting boundaries between a region where $V_A<V_B$ and another where $V_B<V_A$.

We will also discuss the cases of CP conservation. The trivial ones are at boundaries of the CP-violating parameter space. In addition, we find lower-dimensional manifolds of CP conservation (appearing as points in our two-dimensional plots), totally immersed in a region of explicit CP violation.

Our discussion is limited to the scalar sector, but is on the other hand rather general in the sense that we do not commit ourselves to any particular scheme for the Yukawa couplings. 

The paper is organized as follows. In section~\ref{sect:The model} we review the minimal model that allows for explicit as well as spontaneous CP violation. In 
sections~\ref{sect:CP conservation} and \ref{sect:CP violation} we
discuss the conditions for CP conservation and violation,
respectively. In section~\ref{sec:case study} we illustrate our findings with detailed numerical examples, and in section~\ref{sec:disentangling} we discuss the prospects for experimentally establishing CP violation. Section~\ref{Sec:summary} contains a brief summary, one appendix gives explicit minimization conditions, whereas another relates potential parameters to invariants.

\section{The model}
\label{sect:The model}
\setcounter{equation}{0}

The scalar potential of the 2HDM shall be parametrized in the standard fashion:
\begin{align}
V(\Phi_1,\Phi_2) &= -\frac12\left\{m_{11}^2\Phi_1^\dagger\Phi_1
+ m_{22}^2\Phi_2^\dagger\Phi_2 + \left[m_{12}^2 \Phi_1^\dagger \Phi_2
+ \hc\right]\right\} \nonumber \\
& + \frac{\lambda_1}{2}(\Phi_1^\dagger\Phi_1)^2
+ \frac{\lambda_2}{2}(\Phi_2^\dagger\Phi_2)^2
+ \lambda_3(\Phi_1^\dagger\Phi_1)(\Phi_2^\dagger\Phi_2) \nonumber \\
&+ \lambda_4(\Phi_1^\dagger\Phi_2)(\Phi_2^\dagger\Phi_1)
+ \frac12\left[\lambda_5(\Phi_1^\dagger\Phi_2)^2 + \hc\right]\nonumber\\
&+\left\{\left[\lambda_6(\Phi_1^\dagger\Phi_1)+\lambda_7
(\Phi_2^\dagger\Phi_2)\right](\Phi_1^\dagger\Phi_2)
+{\rm \hc}\right\},
\label{Eq:v12}
\end{align}
with
\begin{equation}
\Phi_i=\left(
\begin{array}{c}\varphi_i^+\\ (v_i+\eta_i+i\chi_i)/\sqrt{2}
\end{array}\right), \quad
i=1,2.
\label{Eq:basis}
\end{equation}
All parameters in (\ref{Eq:v12}) are real, except for $m_{12}^2$, $\lambda_5$, $\lambda_6$ and $\lambda_7$, 
which in general could be complex. In the presence of CP violation the neutral sector comprises
3 scalars, $H_i$ ($i=1,2,3$), of undefined CP properties, which are defined through the diagonalization
of the mass-squared matrix, $\mcal^2$, by an orthogonal rotation matrix $R$:
\begin{equation} \label{Eq:R-def}
\begin{pmatrix}
H_1 \\ H_2 \\ H_3
\end{pmatrix}
=R
\begin{pmatrix}
\eta_1 \\ \eta_2 \\ \eta_3
\end{pmatrix},
\end{equation}
satisfying
\begin{equation}
\label{Eq:cal-M}
R{\cal M}^2R^{\rm T}={\cal M}^2_{\rm diag}={\rm diag}(M_1^2,M_2^2,M_3^2),
\end{equation}
and parametrized e.g. in terms of three rotation angles $\alpha_i$ as \cite{Accomando:2006ga}
\begin{equation}     \label{Eq:R-angles}
R
=\begin{pmatrix}
c_1\,c_2 & s_1\,c_2 & s_2 \\
- (c_1\,s_2\,s_3 + s_1\,c_3) 
& c_1\,c_3 - s_1\,s_2\,s_3 & c_2\,s_3 \\
- c_1\,s_2\,c_3 + s_1\,s_3 
& - (c_1\,s_3 + s_1\,s_2\,c_3) & c_2\,c_3
\end{pmatrix}
\end{equation}
with $c_i=\cos\alpha_i$, $s_i=\sin\alpha_i$.
In Eq.~(\ref{Eq:R-def}), $\eta_3 \equiv -\sin\beta\chi_1+\cos\beta\chi_2$
is the combination of $\chi_i$ which is 
orthogonal to the neutral Nambu--Goldstone boson.
Here, $\tan\beta\equiv v_2/v_1$.

We constrain the model by demanding that there exists a basis for $(\Phi_1,\Phi_2)$ in which
the VEVs are real and $\lambda_6=\lambda_7 = 0$. Then the quartic terms of the potential
are invariant under the $\zcal_2$ symmetry $\Phi_i\to \pm \Phi_i$. The symmetry, when
imposed upon the whole Lagrangian (except for the soft-breaking quadratic terms in our potential) eliminates flavour-changing neutral currents (FCNC) 
which otherwise appear in Yukawa interactions. 
We choose to work in this particular basis. 
By choosing another basis, we will in general lose its simplicity by introducing non-zero $\lambda_6$ and $\lambda_7$, and the VEVs may also acquire a phase. 
This will be illustrated by explicit examples later on. This model is the simplest setting in which the 2HDM may give CP violation.
 
We shall also ensure vacuum stability, for that we assume that the potential is positive 
at large field strength irrespective of the direction in the field space.
The positivity conditions for the most general case with $\lambda_6,\lambda_7\neq 0$ (no $\zcal_2$ symmetry) suitable for  a numerical study was formulated in \cite{El Kaffas:2006nt}, and solved in the geometrical approach of \cite{Ivanov:2006}.
Here we limit ourselves to the case with $\lambda_6=\lambda_7=0$,
the positivity conditions then read:
\beq
\lambda_1>0, \lsp \lambda_2>0, \lsp \lambda_3+{\rm min}[0,\lambda_4-|\lambda_5|]
> -\sqrt{\lambda_1 \lambda_2}.
\label{vac_stab}
\eeq

The freedom in choosing a different basis for $(\Phi_1,\Phi_2)$ could be parametrized by the following $U(2)$
transformation\footnote{The parameters of the potential are unaltered by the choice of $\psi$. The transformed VEVs, however, will depend on $\psi$. 
Thus, a suitable choice of $\psi$ allows us to cancel a common phase of the VEVs.}:
\beq
\left(
\begin{array}{c}\bar{\Phi}_1\\ \bar{\Phi}_2
\end{array}\right)
=
e^{i\psi}\left(
\begin{array}{cc}\cos\theta & e^{-i\xi}\sin\theta\\ -e^{i\chi}\sin\theta & e^{i(\chi-\xi)}\cos\theta
\end{array}\right)
\left(
\begin{array}{c}\Phi_1\\ \Phi_2
\end{array}\right).
\label{U(2)}
\eeq

In our analysis the input parameters will be scalar masses $M_{1,2}$, $\mch$, the angles $\alpha_i$ of the neutral-sector 
rotation matrix, and
\begin{equation} \label{Eq:def-lambda}
\mu^2\equiv\frac{v^2}{2v_1v_2}\Re m_{12}^2,
\end{equation} 
along with a $U(1)_\text{em}$-preserving minimum (defining  $\tan\beta$) that is taken to be real. Note that
reality of the VEVs can always be achieved by an appropriate phase rotation of $\Phi_i$ and
therefore does not compromise the generality of our approach. It is easy to see
that the adopted input parameters are
sufficient to determine all the potential parameters\footnote{
When $\lambda_6=\lambda_7=0$,
the potential contains 10 real parameters. Two of the mass parameters could be swapped for
VEVs via the minimization conditions, see Appendix~\ref{Appendix A. Minimum conditions}.  
The third minimization condition eliminates 1 parameter so that we 
eventually get 9 parameters. Those could be determined in terms of 3 masses, 3 mixing angles,
$\mu^2$ and 2 VEVs. For the input masses we use $M_1$, $M_2$ and  $\mch$, then $M_3$
is calculable, see \cite{WahabElKaffas:2007xd} for details. Alternatively, one could take $M_3$ as input rather than the ratio $\tan\beta=v_2/v_1$.}.

In our analysis we will assume that the minimum specified by $v_{1,2}$ satifies the constraint 
$v_1^2+v_2^2\sim (246\gev)^2$. However it may happen that this minimum is not the global minimum
(vacuum), so we will use the subscript $A$ for our starting minimum to distinguish it from 
other minima we encounter. Thus,
\beq
\langle \Phi_1 \rangle_A=\frac{1}{\sqrt{2}}\left(
\begin{array}{c}0\\
v_1
\end{array}\right),\quad
\langle \Phi_2 \rangle_A =\frac{1}{\sqrt{2}}\left(
\begin{array}{c}0\\
v_2
\end{array}\right).
\eeq

In this paper we are going to study the CP-properties of the model with particular emphasis on distinguishing explicit and spontaneous CP violation. Necessary and sufficient criteria for how
to distinguish these two types of CP violation has been worked out by different groups. In \cite{Davidson:2005,Gunion:2005ja} a tensorial approach has been used for this purpose, while in 
\cite{Ivanov:2005,Ivanov:2006,Maniatis:2007,Nishi:2006} geometric methods have been developed 
for the same purpose. In our work, we ``control" the vacuum
since we start with a set of physical masses and  the location of the vacuum as input parameters. The parameters of the potential are determined from our set of (physical) input parameters.
We have found the approach of \cite{Davidson:2005,Gunion:2005ja} more convenient for our purposes, and thus we have adopted their tensorial approach. However, we have verified that for the model
which was considered in this paper, the conditions for
CP conservation obtained in \cite{Ivanov:2005,Ivanov:2006,Maniatis:2007,Nishi:2006}
coincides with those found in \cite{Davidson:2005,Gunion:2005ja}.

Studying the CP properties of the model, we will sometimes need to express the parameters of the potential also in a 
different basis. By changing basis, we will in these cases see the true nature of CP in our model. Any two different bases are related by a $U(2)$-transformation (\ref{U(2)}). 
In particular, we shall be interested in the cases where a basis exists in which all the parameters of the 
potential are real \cite{Davidson:2005,Gunion:2005ja}. This is possible for the cases where CP is conserved or broken spontaneously. We will use a bar-notation to distinguish the parameters of the potential and the fields in this basis, 
i.e., $\bar\lambda_i$, $\bar m_{ij}$ and $\bar\Phi_i$ from the parameters we originally started from.

We shall limit ourselves in this study to a model defined by imposing the $\zcal_2$ symmetry for dimension-4 operators in the Lagrangian formulated in a certain initial basis. Then, in this basis, $\lambda_6=\lambda_7=0$ and tree-level Flavour-Changing Neutral Currents are absent in Yukawa couplings~\cite{Glashow:1976nt}. 
This symmetry will be softly violated by a dimension-2 operator $\Phi_1^\dagger \Phi_2$, here referred to as the $m_{12}^2$~term.
Note however, that any $U(2)$ rotation would in general reintroduce non-zero $\lambda_6$ and $\lambda_7$.
In particular, it is worth noticing that a rotation could be adopted to eliminate the $m_{12}^2$ term. That
would introduce $\lambda_6$- and $\lambda_7$-terms, so that the $\zcal_2$ would appear hardly broken in the other basis. 
However, the coefficients of those terms would be correlated in such a way that the
renormalizability would be preserved exactly in the same manner as in the initial basis containing soft breaking through 
non-zero $m_{12}^2$ with vanishing $\lambda_6$ and $\lambda_7$.

We shall throughout this paper have repeated need for the phases of $m_{12}^2$ and $\lambda_5$, 
so we introduce the following notation for this purpose,
\begin{equation} \label{Eq:parameter-phases}
m_{12}^2=|m_{12}^2| e^{i\alpha}, \quad
\lambda_5=|\lambda_5|e^{i\gamma},\quad 0\leq\alpha,\gamma<2\pi.
\end{equation}

If CP is conserved, or spontaneously violated, then a basis exists in which all the parameters of the potential are real. Thus, in this basis the potential (\ref{Eq:v12}) can be written as
\begin{align}
\bar V(\bar\Phi_1,\bar\Phi_2) &= -\frac12\left\{\bar m_{11}^2\bar\Phi_1^\dagger\bar\Phi_1
+ \bar m_{22}^2\bar\Phi_2^\dagger\bar\Phi_2 + \bar m_{12}^2 \left[\bar\Phi_1^\dagger \bar\Phi_2
+ \hc\right]\right\} \nonumber \\
& + \frac{\bar\lambda_1}{2}(\bar\Phi_1^\dagger\bar\Phi_1)^2
+ \frac{\bar\lambda_2}{2}(\bar\Phi_2^\dagger\bar\Phi_2)^2
+ \bar\lambda_3(\bar\Phi_1^\dagger\bar\Phi_1)(\bar\Phi_2^\dagger\bar\Phi_2) \nonumber \\
&+ \bar\lambda_4(\bar\Phi_1^\dagger\bar\Phi_2)(\bar\Phi_2^\dagger\bar\Phi_1)
+ \frac12\bar\lambda_5\left[(\bar\Phi_1^\dagger\bar\Phi_2)^2 + \hc\right]\nonumber\\
&+\left[\bar\lambda_6(\bar\Phi_1^\dagger\bar\Phi_1)
+\bar\lambda_7
(\bar\Phi_2^\dagger\bar\Phi_2)\right]\left[(\bar\Phi_1^\dagger\bar\Phi_2)
+{\rm \hc}\right],\label{v12realbasis}
\end{align}
where now all the $\bar{\lambda_i}$ and $\bar{m}_{ij}^2$ are real. This basis has the property that if CP is conserved, both VEVs are real, while if CP is spontaneously violated, the VEV of one doublet is complex. Our starting minimum ``A'' will in this basis be denoted $(\langle\bar\Phi_{1}\rangle_A,\langle\bar\Phi_{2}\rangle_A)$.

\section{CP conservation}
\label{sect:CP conservation}
\setcounter{equation}{0}

In any 2HDM, CP is conserved if and only if the three invariants $J_1$, $J_2$ 
and $J_3$ \cite{Lavoura:1994fv,Davidson:2005,Gunion:2005ja} are all real.
In a model in which $\lambda_6=\lambda_7=0$ and the VEVs are real, 
these invariants can be written in a compact form \cite{Grzadkowski:2009bt}:
\begin{eqnarray} \label{eq:im_J1}
\Im J_1&=&-\frac{2}{v^2}\Im\bigl[\hat{v}_{\bar{a}}^* Y_{a\bar{b}} Z_{b\bar{d}}^{(1)}\hat{v}_d\bigr] \nonumber\\
&=&-\frac{v_1^2v_2^2}{v^4}(\lambda_1-\lambda_2)\Im \lambda_5\\
\Im J_2&=&\frac{2}{v^4}\Im\bigl[\hat{v}_{\bar{b}}^* \hat{v}_{\bar{c}}^* Y_{b\bar{e}} Y_{c\bar{f}} Z_{e\bar{a}f\bar{d}}\hat{v}_a\hat{v}_d\bigr] \nonumber\\
&=&-\frac{v_1^2v_2^2}{v^8}
\left[\left((\lambda_1-\lambda_3-\lambda_4)^2-|\lambda_5|^2\right) v_1^4
+2(\lambda_1-\lambda_2) \Re \lambda_5 v_1^2v_2^2\right.\nonumber\\
&&\hspace*{1.5cm}\left.
-\left((\lambda_2-\lambda_3-\lambda_4)^2-|\lambda_5|^2\right) v_2^4\right]
\Im \lambda_5\\
\Im J_3&=&\Im\bigl[\hat{v}_{\bar{b}}^* \hat{v}_{\bar{c}}^* Z_{b\bar{e}}^{(1)} Z_{c\bar{f}}^{(1)}Z_{e\bar{a}f\bar{d}}\hat{v}_a\hat{v}_d\bigr] \nonumber\\
&=&\frac{v_1^2v_2^2}{v^4}(\lambda_1-\lambda_2)
(\lambda_1+\lambda_2+2\lambda_4)\Im \lambda_5
 \label{eq:im_J3}
\end{eqnarray}
The first line of each of these three equations defines the invariant \cite{Davidson:2005,Gunion:2005ja} (see also \cite{Lavoura:1994fv,Botella:1994cs}), whereas the second line is the model-specific expression for the invariant written out in our starting basis.
It is worth noting the absence of $\Im m_{12}^2$ above, its presence is hidden
since the minimization condition (\ref{Eq:lambda5-m12_sq}) has been invoked to express
$\Im m_{12}^2$ through $\Im \lambda_5$.

Thus, CP conservation requires
\begin{equation}
\Im J_1=\Im J_2=\Im J_3=0.
\end{equation}
The conditions under which CP is conserved in such a model are described in 
\cite{Grzadkowski:2009bt}. They are labeled CPC1 to CPC5, and defined by
\begin{itemize}
\item CPC1: $v_1= 0$
\item CPC2: $v_2= 0$
\item CPC3: $\Im\lambda_5=0$
\item CPC4: $\lambda_1=\lambda_2$ and $v_1= v_2$
\item CPC5: $\lambda_1=\lambda_2$ and $(\lambda_1-\lambda_3-\lambda_4)^2=|\lambda_5|^2$
\end{itemize}
While CPC1--CPC3 are quite trivial it is worth paying some attention to the two remaining conditions. Both require {\it two} conditions to be satisfied, and will thus only be satisfied in a lower-dimensional parameter space, as compared with the former three cases.
\subsection{CPC4: $\lambda_1=\lambda_2$ and $v_1= v_2$}
It can be shown that in this case the following $U(2)$ transformation will make the parameters of the potential and the VEVs simultaneously real:
\beq
\left(
\begin{array}{c}\bar{\Phi}_1\\ \bar{\Phi}_2
\end{array}\right)
=
e^{i\psi}\left(
\begin{array}{cc}\cos\frac{\pi}{4} & e^{-i\xi}\sin\frac{\pi}{4}\\ i\sin\frac{\pi}{4} & -i e^{-i\xi}\cos\frac{\pi}{4}
\end{array}\right)
\left(
\begin{array}{c}\Phi_1\\ \Phi_2
\end{array}\right)
\eeq
where $\xi=-\gamma/2$, $\psi=-\gamma/4$ and $\gamma=\arg (\lambda_5)$.

We find that after this transformation
\begin{eqnarray}
\bar{m}_{12}^2 &=& \left[\Re m_{12}^2-2|\lambda_5|v_1^2\cos^2\frac{\gamma}{2}\right]\sin\frac{\gamma}{2},\nonumber\\
\bar{\lambda}_5 &=& -\frac{1}{2}(\lambda_1-\lambda_3-\lambda_4+|\lambda_5|),\nonumber\\
\bar{\lambda}_6 &=& 0,\nonumber\\
\bar{\lambda}_7 &=& 0, \nonumber\\
\bar{\lambda}_1&=&\bar{\lambda}_2.
\end{eqnarray}
Furthermore,
\begin{eqnarray}
\langle \bar{\Phi}_1 \rangle_A&=&\left(
\begin{array}{c}0\\
v_1\cos\frac{\gamma}{4}
\end{array}\right)\\
\langle \bar{\Phi}_2 \rangle_A &=&\left(
\begin{array}{c}0\\
v_1\sin\frac{\gamma}{4}
\end{array}\right)
\end{eqnarray}
with
\begin{equation}
\tan\bar\beta=\tan\frac{\gamma}{4}.
\end{equation}

\subsection{CPC5: $\lambda_1=\lambda_2$ and $(\lambda_1-\lambda_3-\lambda_4)^2=|\lambda_5|^2$}

Let us consider two different cases for which this can happen:
\begin{itemize}
\item
Case 1: $\lambda_1=\lambda_2$ and $\lambda_1-\lambda_3-\lambda_4=-|\lambda_5|$
\item
Case 2: $\lambda_1=\lambda_2$ and $\lambda_1-\lambda_3-\lambda_4=+|\lambda_5|$
\end{itemize}
In both these cases a basis exists in which all the parameters of the potential and the VEVs are simultaneously real.
\subsubsection{Case 1: $\lambda_1=\lambda_2$ and $\lambda_1-\lambda_3-\lambda_4=-|\lambda_5|$}
In this case, when $v_1+v_2\cos(\gamma/2)\neq0$ the following $U(2)$ transformation will make all the parameters of
the potential and the VEVs real:
\begin{align}
\left(
\begin{array}{c}\bar{\Phi}_1\\ \bar{\Phi}_2
\end{array}\right)
&={\rm sgn}(v_1+v_2\cos\textstyle\frac{\gamma}{2}) \nonumber \\
&\times
e^{i\psi}\left(
\begin{array}{cc}\cos\frac{\pi}{4} & e^{-i\xi}\sin\frac{\pi}{4}\\ 
-{\rm sgn}(v_2-v_1)e^{i\chi}\sin\frac{\pi}{4} & {\rm sgn}(v_2-v_1)e^{i(\chi-\xi)}\cos\frac{\pi}{4}
\end{array}\right)
\left(
\begin{array}{c}\Phi_1\\ \Phi_2
\end{array}\right)
\end{align}
where 
\begin{equation}
\xi=-\frac{\gamma}{2},\quad 
\chi=\arctan\frac{2v_1 v_2 \sin\frac{\gamma}{2}}{v_1^2-v_2^2},\quad
\psi=-\arctan\frac{v_2 \sin\frac{\gamma}{2}}{v_1+v_2\cos\frac{\gamma}{2}}
\end{equation}
and $\gamma=\arg (\lambda_5)$.

After this transformation we have
\begin{eqnarray}
\bar{m}_{12}^2 &=& \frac{\left[\Re m_{12}^2-2|\lambda_5|v_1 v_2 \cos^2\textstyle\frac{\gamma}{2} \right]\sqrt{v_1^4+v_2^4-2v_1^2 v_2^2 \cos\gamma}}{2 v_1 v_2},\nonumber\\
\bar{\lambda}_5 &=& 0,\nonumber\\
\bar{\lambda}_6 &=& 0,\nonumber\\
\bar{\lambda}_7 &=& 0.
\end{eqnarray}
Furthermore,
\begin{equation}
\bar{\lambda}_1=\bar{\lambda}_2,
\end{equation}
and the transformed minimum becomes
\begin{eqnarray}
\langle \bar{\Phi}_1 \rangle_A&=&\frac{1}{2}\left(
\begin{array}{c}0\\
\sqrt{v_1^2+v_2^2+2v_1 v_2\cos\textstyle\frac{\gamma}{2}}
\end{array}\right)\\
\langle \bar{\Phi}_2 \rangle_A &=&\frac{1}{2}\left(
\begin{array}{c}0\\
\sqrt{v_1^2+v_2^2-2v_1 v_2\cos\textstyle\frac{\gamma}{2}}
\end{array}\right)
\end{eqnarray}
meaning the VEVs are all real. This corresponds to CP conservation.
However, the value of $\tan\beta$ has also been transformed,
\begin{equation}
\tan\bar\beta=\sqrt{\frac{1+\tan^2\beta-2\tan\beta\cos\frac{\gamma}{2}}{1+\tan^2\beta+2\tan\beta\cos\frac{\gamma}{2}}}.
\end{equation}

Finally, considering the special case when $v_1+v_2\cos\textstyle\frac{\gamma}{2}=0$ (which could occur for $\tan\beta>1$), we need to use $\psi=-\textstyle\frac{\pi}{2}$ in the above $U(2)$ transformation
in order to make the parameters and the VEVs real. The transformed quantities now become
\begin{eqnarray}
\bar{m}_{12}^2 &=& \frac{(\Re m_{12}^2 v_2-2|\lambda_5|v_1^3)\sqrt{(v_2^2+3v_1^2)(v_2^2-v_1^2)}}{2 v_1 v_2^2},\nonumber\\
\bar{\lambda}_5 &=& 0,\nonumber\\
\bar{\lambda}_6 &=& 0,\nonumber\\
\bar{\lambda}_7 &=& 0,
\end{eqnarray}
and the transformed minimum is given by
\begin{eqnarray}
\langle \bar{\Phi}_1 \rangle_A&=&\frac{1}{2}\left(
\begin{array}{c}0\\
\sqrt{v_2^2-v_1^2}
\end{array}\right)\\
\langle \bar{\Phi}_2 \rangle_A &=&\frac{1}{2}\left(
\begin{array}{c}0\\
\sqrt{v_2^2+3v_1^2}
\end{array}\right)
\end{eqnarray}
with
\begin{equation} \label{Eq:case-1-tanbetabar}
\tan\bar\beta=\sqrt{\frac{\tan^2\beta+3}{\tan^2\beta-1}}.
\end{equation}

\subsubsection{Case 2: $\lambda_1=\lambda_2$ and $\lambda_1-\lambda_3-\lambda_4=+|\lambda_5|$}
In this case, when $v_1+v_2\sin(\gamma/2)\neq0$ the following $U(2)$ transformation will make all the parameters of
the potential and the VEVs real:
\begin{align}
\left(
\begin{array}{c}\bar{\Phi}_1\\ \bar{\Phi}_2
\end{array}\right)
&={\rm sgn}(v_1+v_2\sin\textstyle\frac{\gamma}{2}) \nonumber \\
&\times e^{i\psi}\left(
\begin{array}{cc}\cos\frac{\pi}{4} & e^{-i\xi}\sin\frac{\pi}{4}\\ 
-{\rm sgn}(v_2-v_1)e^{i\chi}\sin\frac{\pi}{4} & {\rm sgn}(v_2-v_1)e^{i(\chi-\xi)}\cos\frac{\pi}{4}
\end{array}\right)
\left(
\begin{array}{c}\Phi_1\\ \Phi_2
\end{array}\right)
\end{align}
where
\begin{equation}
\xi=\frac{\pi}{2}-\frac{\gamma}{2}, \quad
\chi=-\arctan\frac{2v_1 v_2 \cos\frac{\gamma}{2}}{v_1^2-v_2^2}, \quad
\psi=\arctan\frac{v_2 \cos\frac{\gamma}{2}}{v_1+v_2\sin\frac{\gamma}{2}}
\end{equation}
and $\gamma=\arg (\lambda_5)$.

We find that after this transformation
\begin{eqnarray}
\bar{m}_{12}^2 &=& \frac{\left[\Re m_{12}^2+2|\lambda_5|v_1 v_2\sin^2\textstyle\frac{\gamma}{2} \right]\sqrt{v_1^4+v_2^4+2v_1^2 v_2^2 \cos\gamma}}{2 v_1 v_2},\nonumber\\
\bar{\lambda}_5 &=& 0,\nonumber\\
\bar{\lambda}_6 &=& 0,\nonumber\\
\bar{\lambda}_7 &=& 0.
\end{eqnarray}
Furthermore,
\begin{eqnarray}
\langle \bar{\Phi}_1 \rangle_A &=&\frac{1}{2}\left(
\begin{array}{c}0\\
\sqrt{v_1^2+v_2^2+2v_1 v_2\sin\textstyle\frac{\gamma}{2}}
\end{array}\right)\\
\langle \bar{\Phi}_2 \rangle_A &=&\frac{1}{2}\left(
\begin{array}{c}0\\
\sqrt{v_1^2+v_2^2-2v_1 v_2\sin\textstyle\frac{\gamma}{2}}
\end{array}\right)
\end{eqnarray}
meaning they are all real. This corresponds to CP conservation. Furthermore,
\begin{equation}
\tan\bar\beta=\sqrt{\frac{1+\tan^2\beta-2\tan\beta\sin\frac{\gamma}{2}}{1+\tan^2\beta+2\tan\beta\sin\frac{\gamma}{2}}}.
\end{equation}

Finally, considering the special case when $v_1+v_2\sin\textstyle\frac{\gamma}{2}=0$, we have to 
use $\psi=\textstyle\frac{\pi}{2}$ in the above $U(2)$ transformation
in order to make the parameters and the VEVs real. The transformed quantities now become
\begin{eqnarray}
\bar{m}_{12}^2 &=& \frac{(\Re m_{12}^2 v_2+2|\lambda_5|v_1^3)\sqrt{(v_2^2+3v_1^2)(v_2^2-v_1^2)}}{2 v_1 v_2^2},\nonumber\\
\bar{\lambda}_5 &=& 0,\nonumber\\
\bar{\lambda}_6 &=& 0,\nonumber\\
\bar{\lambda}_7 &=& 0,
\end{eqnarray}
and the transformed minimum is given by
\begin{eqnarray}
\langle \bar{\Phi}_1 \rangle_A&=&\frac{1}{2}\left(
\begin{array}{c}0\\
\sqrt{v_2^2-v_1^2}
\end{array}\right)\\
\langle \bar{\Phi}_2 \rangle_A &=&\frac{1}{2}\left(
\begin{array}{c}0\\
\sqrt{v_2^2+3v_1^2}
\end{array}\right)
\end{eqnarray}
and $\tan\bar\beta$ by Eq.~(\ref{Eq:case-1-tanbetabar}).

We note that in both these cases CPC4 and CPC5 (and their subcases), $\bar\lambda_6$ and $\bar\lambda_7$ remain zero, but $\tan\beta$ is transformed into a different value $\tan\bar\beta$.
\section{CP violation}
\label{sect:CP violation}
\setcounter{equation}{0}

In any 2HDM, CP is conserved if and only if the three invariants $J_1$, $J_2$ 
and $J_3$ \cite{Davidson:2005,Gunion:2005ja} are all real, see Eqs.~(\ref{eq:im_J1})--(\ref{eq:im_J3}).

Thus, CP violation requires
\begin{equation}
\Im J_1\neq0 \quad\text{\it and/or}\quad
\Im J_2\neq0 \quad\text{\it and/or}\quad
\Im J_3\neq0.
\end{equation}

\subsection{Explicit CP violation}
According to \cite{Davidson:2005,Gunion:2005ja}, we have to check four invariant quantities,
$I_{Y3Z}$, $I_{2Y2Z}$, $I_{3Y3Z}$ and $I_{6Z}$ to determine whether CP is broken 
spontaneously or explicitly in a CP-violating model. In any 2HDM, CP is broken
explicitly if at least one of these invariants is non-zero. 
This means that there exists no basis for which all the parameters
of the potential are real. 

In a 2HDM with $\lambda_6=\lambda_7=0$, and with real VEVs, two 
of these invariants are zero, and the other two can be written in a compact form:
\begin{eqnarray}
I_{Y3Z}&=&\Im \bigl[Z_{a\bar{c}}^{(1)} Z_{e\bar{b}}^{(1)} Z_{b\bar{e}c\bar{d}}Y_{d\bar{a}}\bigr] \nonumber\\
&=&0,\\
I_{2Y2Z}&=&\Im \bigl[Y_{a\bar{b}}  Y_{c\bar{d}} Z_{b\bar{a}d\bar{f}} Z_{f\bar{c}}^{(1)}\bigr] \nonumber\\
&=&\frac{1}{4}(\lambda_1-\lambda_2)\Im\left[(m_{12}^2)^2\lambda_5^*\right]\nonumber\\
&=&\frac{v_1^2v_2^2}{4v^4}(\lambda_1-\lambda_2)
\left[4v^2\mu^2\Re\lambda_5-4\mu^4+v^4(\Im\lambda_5)^2\right]\Im\lambda_5,\\
I_{3Y3Z}&=&\Im \bigl[Z_{a\bar{c}b\bar{d}} Z_{c\bar{e}d\bar{g}} Z_{e\bar{h}f\bar{q}} Y_{g\bar{a}} Y_{h\bar{b}} Y_{q\bar{f}}\bigr]\nonumber\\
&=&-\frac{1}{8}(m_{11}^2-m_{22}^2)
\left[(\lambda_1-\lambda_3-\lambda_4)(\lambda_2-\lambda_3-\lambda_4)-|\lambda_5|^2\right]
\Im\left[(m_{12}^2)^2\lambda_5^*\right]\nonumber\\
&=&-\frac{v_1^2v_2^2}{8v^6}
\left[(\lambda_1-\lambda_3-\lambda_4)(\lambda_2-\lambda_3-\lambda_4)-|\lambda_5|^2\right]\nonumber\\
&&\times
\left[(v_1^2-v_2^2)(2\mu^2-v^2(\lambda_3+\lambda_4+\Re\lambda_5))+v^2(v_1^2\lambda_1-v_2^2\lambda_2)\right]\nonumber\\
&&\times
\left[4v^2\mu^2\Re\lambda_5-4\mu^4+v^4(\Im\lambda_5)^2\right]\Im\lambda_5,\\
I_{6Z}&=&\Im \bigl[Z_{a\bar{b}c\bar{d}}Z_{b\bar{f}}^{(1)} Z_{d\bar{h}}^{(1)} Z_{f\bar{a}j\bar{k}}Z_{k\bar{j}m\bar{n}}Z_{n\bar{m}h\bar{c}}\bigr] \nonumber\\
&=&0.
\end{eqnarray}

Some comments are here in order:
\begin{trivlist}
\item[$-$]
The first line of each of these equations is the definition of the invariant \cite{Davidson:2005,Gunion:2005ja}.
\item[$-$]
The second line is the model-specific expression of the invariant given in our starting basis before applying the minimization conditions.
\item[$-$]
In order to obtain the third form for $I_{2Y2Z}$ we have used the relation (\ref{Eq:def-lambda})  defining $\mu^2$, and (\ref{Eq:lambda5-m12_sq}) between $\Im m_{12}^2$ and $\Im\lambda_5$, obtained by minimization of the potential for real VEVs.
\item[$-$]
In order to obtain the third form for $I_{3Y3Z}$ we have expressed $m_{11}^2$ and $m_{22}^2$ in terms of the $\lambda$s, according to the 
minimization conditions (\ref{Eq:min-cond-1}) and (\ref{Eq:min-cond-2}).
\end{trivlist}

In general the CP violation is {\it explicit} if
\begin{equation}
I_{Y3Z}\neq0 \quad\text{\it and/or}\quad
I_{2Y2Z}\neq0 \quad\text{\it and/or}\quad
I_{3Y3Z}\neq0 \quad\text{\it and/or}\quad
I_{6Z}\neq0.
\end{equation}
However in the simple model defined by Eq.~(\ref{Eq:v12}), the non-trivial part of this is
\begin{equation}
I_{2Y2Z}\neq0 \quad\text{\it and/or}\quad
I_{3Y3Z}\neq0.
\end{equation}
\subsection{Spontaneous CP violation}
In the case when
\begin{equation}
I_{Y3Z}=
I_{2Y2Z}=
I_{3Y3Z}=
I_{6Z}=0,
\end{equation}
CP is either conserved or broken spontaneously. If, in addition, at least one of 
the $J_i$ is complex, the CP violation is spontaneous. This means that there
exists a choice of basis where all the parameters of the potential are real, 
but then the vacuum breaks CP (complex VEVs).

For CP to be broken {\it spontaneously} it is necessary that the following five
conditions are satisfied simultaneously (failure to do so means the model is 
CP conserving):
\begin{itemize}
\item $v_1\neq 0$
\item $v_2\neq 0$
\item $\Im\lambda_5\neq0$
\item $\lambda_1\neq\lambda_2$ or $v_1\neq v_2$
\item $\lambda_1\neq\lambda_2$ or $(\lambda_1-\lambda_3-\lambda_4)^2\neq|\lambda_5|^2$
\end{itemize}
In addition, one or both of the following conditions emerging from the requirement 
that $I_{2Y2Z}=0$ and $I_{3Y3Z}=0$ must be satisfied (otherwise the
CP violation would be explicit):
\begin{itemize}
\item SCPV1:
\beq
4\frac{\mu^2}{v^2}\Re\lambda_5-4\left(\frac{\mu^2}{v^2}\right)^2+(\Im\lambda_5)^2=0 \quad
({\rm or~equivalently} \quad
\Im\left[(m_{12}^2)^2\lambda_5^*\right]=0) \label{scpv1}
\eeq
\item SCPV2: 
\beq
\lambda_1=\lambda_2,\lsp \lambda_1=\lambda_3+\lambda_4+\Re\lambda_5-2\frac{\mu^2}{v^2} \quad
({\rm or~equivalently} \quad
\lambda_1=\lambda_2, \;  m_{11}^2=m_{22}^2)
\label{scpv2}
\eeq
\end{itemize}
Note that these conditions refer to the basis defined by Eq.~(\ref{Eq:v12}). The above conditions ensure that
the potential is indeed CP invariant, and CP is only broken by the VEVs.

An important comment is here in order. Assuming that $U(1)_\text{em}$ is not spontaneously broken, we can, 
without compromising generality, assume that in any basis $\langle\bar\Phi_1\rangle$ is real while 
$\langle\bar\Phi_2\rangle$ is complex. The value of the potential at the minimum will be
$V_{\rm min}=\bar V(\langle\bar\Phi_1\rangle_A,\langle\bar\Phi_2\rangle_A)$. 
Complex conjugating both sides of (\ref{v12realbasis}) it is easy to see that 
\begin{eqnarray}
V_{\rm min}=\bar V(\langle\bar\Phi_1\rangle_A,\langle\bar\Phi_2\rangle_A)=\bar V(\langle\bar\Phi_1\rangle_A,\langle\bar\Phi_2\rangle_A^*).
\end{eqnarray}
This means that there exists another minimum of exactly the same depth as our starting minimum A. In the real basis, this second minimum is located at a position in 
$\Phi_i$-space that is the complex conjugate of the location of minimum A. Let us label this second minimum B. Thus
\begin{eqnarray}
\langle\bar\Phi_1\rangle_B=\langle\bar\Phi_1\rangle_A,\hspace*{1cm}\langle\bar\Phi_2\rangle_B=\langle\bar\Phi_2\rangle_A^*.
\end{eqnarray}
Thus, when we have SCPV, 
{\it there exist two minima of the same depth which (in the real basis) are complex conjugates of each other.}
\subsubsection{SCPV1: $\Im\left[(m_{12}^2)^2\lambda_5^*\right]=0$}

Invoking the definitions (\ref{Eq:parameter-phases}), the condition $\Im\left[(m_{12}^2)^2\lambda_5^*\right]=0$ becomes:
\beq
|m_{12}^2|^2|\lambda_5|\Im(e^{i(2\alpha-\gamma)})=0,
\eeq
which is satisfied when $\sin(2\alpha-\gamma)=0$. This in turns means that $\cos(2\alpha-\gamma)=\pm 1$, or
\begin{equation}
\gamma=2\alpha+n\pi, \quad n\text{ integer}.
\end{equation}

In this case, for $\lambda_6=\lambda_7=0$, the following U(2) transformation will make all the parameters of
the potential real:
\beq
\left(
\begin{array}{c}\bar{\Phi}_1\\ \bar{\Phi}_2
\end{array}\right)
=
\left(
\begin{array}{cc}1 & 0\\ 0 & e^{i\alpha}
\end{array}\right)
\left(
\begin{array}{c}\Phi_1\\ \Phi_2
\end{array}\right)
\eeq
This transformation yields
\begin{eqnarray}
\bar{m}_{12}^2 &=& m_{12}^2e^{-i\alpha}=|m_{12}^2|,\nonumber\\
\bar{\lambda}_5 &=& \lambda_5 e^{-2i\alpha} = |\lambda_5|e^{-i(2\alpha-\gamma)}=|\lambda_5|\cos(2\alpha-\gamma)=\pm |\lambda_5|,\nonumber\\
\bar{\lambda}_6 &=& 0,\nonumber\\
\bar{\lambda}_7 &=& 0,
\end{eqnarray}
meaning they are all real. This corresponds to spontaneous CP violation. The transformed starting minimum is in this case:

\begin{eqnarray}
\langle \bar{\Phi}_1 \rangle_A &=&\frac{1}{\sqrt{2}}\left(
\begin{array}{c}0\\
v_1
\end{array}\right)\\
\langle \bar{\Phi}_2 \rangle_A &=&\frac{e^{i\alpha}}{\sqrt{2}}\left(
\begin{array}{c}0\\
v_2
\end{array}\right)
\end{eqnarray}

\subsubsection{SCPV2: $\lambda_1=\lambda_2$ and $m_{11}^2=m_{22}^2$}

In this case, the following U(2) transformation will make all the parameters of
the potential real:
\beq
\left(
\begin{array}{c}\bar{\Phi}_1\\ \bar{\Phi}_2
\end{array}\right)
=
\left(
\begin{array}{cc}\cos\frac{\pi}{4} & \sin\frac{\pi}{4}\\ -i\sin\frac{\pi}{4} & i\cos\frac{\pi}{4}
\end{array}\right)
\left(
\begin{array}{c}\Phi_1\\ \Phi_2
\end{array}\right)
\eeq
This transformation yields
\begin{eqnarray}
\bar{m}_{12}^2 &=& \Im (m_{12}^2),\nonumber\\
\bar{\lambda}_5 &=& -\frac{1}{4}(\lambda_1+\lambda_2)
+\frac{1}{2}(\lambda_3+\lambda_4-\Re \lambda_5),\nonumber\\
\bar{\lambda}_6 &=& \frac{1}{2}\Im\lambda_5,\nonumber\\
\bar{\lambda}_7 &=& -\frac{1}{2}\Im\lambda_5,
\end{eqnarray}
meaning they are all real. This corresponds to spontaneous CP violation. The transformed starting minimum is in this case:

\begin{eqnarray}
\langle \bar{\Phi}_1 \rangle_A &=&\frac{1}{2}\left(
\begin{array}{c}0\\
v_1+v_2
\end{array}\right)\\
\langle \bar{\Phi}_2 \rangle_A &=&\frac{i}{2}\left(
\begin{array}{c}0\\
v_2-v_1
\end{array}\right)
\end{eqnarray}

\section{Case studies}
\setcounter{equation}{0}
\label{sec:case study}

We will discuss regions in the parameter space of the model limiting ourselves to the following representative cases:
\ben
\item
$M_1=125\gev$, $M_2= 200\gev$, $\mch=350\gev$, $\mu=250\gev$,  $\tanb=0.5, 1, 2$,
\item
$M_1=125\gev$, $M_2= 200\gev$, $\mch=350\gev$, $\mu=250\gev$,  $\tanb=5, 10, 30$,
\item
$M_1=125\gev$, $M_2= 300\gev$, $\mch=500\gev$, $\mu=300\gev$,  $\tanb=0.5, 1, 2$,
\item
$M_1=125\gev$, $M_2= 300\gev$, $\mch=500\gev$, $\mu=300\gev$,  $\tanb=5, 10, 30$.
\een
For these choices we fix $\alpha_1$ and search through the $(\alpha_2,\alpha_3)$
plane in order to determine regions that are consistent with CP conservation and/or CP violation (explicit or spontaneous). 

\begin{figure}[htb]
\centering
\includegraphics[width=\textwidth]{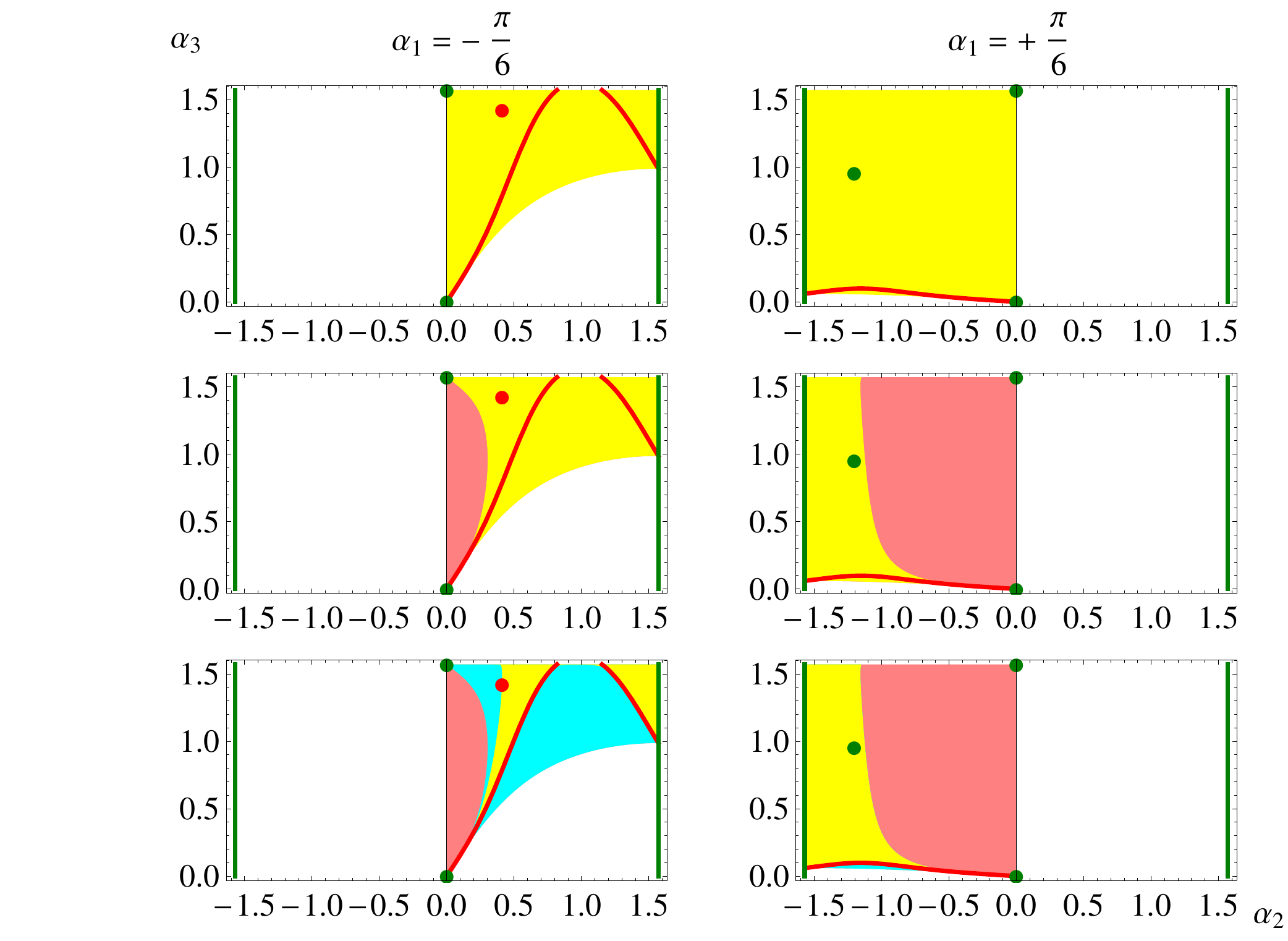}
\caption{For $\tan\beta=2$, and two values of $\alpha_1$ (left: $\alpha_1=-\pi/6$, right: $\alpha_1=+\pi/6$), the top panels show the allowed regions (yellow) in the $\alpha_2$--$\alpha_3$ space after imposing the constraint $M_3>M_2$. Red curves correspond to parameters that satisfy the condition (\ref{scpv1}), while red dots satisfy the condition (\ref{scpv2}). 
Both of these indicate spontaneous CP violation.
Green lines and dots indicate locations of CP conservation.
Middle panels: the positivity constraint (\ref{vac_stab}) is also imposed (pink region disallowed). 
Bottom panels: additionally, the global minimum constraint is imposed (cyan region disallowed).}
\label{fig:singlepanel}
\end{figure}

We start with Fig.~\ref{fig:singlepanel} where, for $\tanb=2$ and $\alpha_1=\pm\pi/6$ it is illustrated
how the different constraints reduce the allowed region of the $(\alpha_2,\alpha_3)$ parameter space. 
The rotation angles are defined according to the conventions of \cite{Accomando:2006ga}, so that
the allowed ranges are $-\pi/2<\alpha_2\leq\pi/2$ and $0\leq\alpha_3\leq\pi/2$. 
It is worth noticing that for given values of $\tan\beta$ and $\alpha_1$, only one of these two quadrants is accessible by allowed models \cite{Khater:2003wq}. (At the border, for $\alpha_2=0$, we have $M_3=M_2$.)

The boundaries of the yellow regions will be of particular interest in the following discussion.
Green lines and dots indicate locations where CP is conserved. Everywhere else, CP is violated.
Red curves and dots indicate where the CP violation is {\it spontaneous.}

In the upper panels of Fig.~\ref{fig:singlepanel}, the yellow region indicates where a consistent solution for $M_3$ (real, 
and satisfying $M_3\geq M_2$) can be found, otherwise white color is adopted. 
In the middle panels, positivity (\ref{vac_stab}) has been imposed. The pink region indicates where positivity is violated. 
In the bottom panels, we also impose the constraint that the starting minimum A shall be global. 
The region forbidden by this constraint is shown in cyan.

As illustrated by the middle and lower panels of Fig.~\ref{fig:singlepanel}, 
there are two kinds of borders which are relevant for the model: 
(i) the border between a region where positivity is satisfied, and where it is not 
(illustrated by yellow and pink in the middle
and bottom panels), and 
(ii) the border between the region where the starting minimum is the global one, 
and where it is not (illustrated by yellow and cyan in the bottom panels). 
We shall refer to these regions as ``physical'' (yellow), ``non-positive'' 
(pink) and ``non-global'' (cyan).

More results are shown in Figs.~\ref{fig:panels-rainbow1}-\ref{fig:panels-rainbow4}.

\subsection{CPC}
Regions of CPC are denoted by green color, they correspond to
parameters for which one of the conditions CPC1--CPC5 specified in 
section~\ref{sect:CP conservation} is satisfied. It is worth 
noting which cases can be realized for our parameter choices.
The trivial cases CPC1 
and CPC2 are not illustrated in our plots. Since we consider only
non-degenerate scalar masses, the case CPV3, i.e.
$\Im{\lambda_5}=0$ corresponds to \cite{ElKaffas:2007rq}: 
\bit
\item $\alpha_2=\pm\pi/2$ (then $R_{11}=R_{12}=R_{23}=R_{33}=0$ and $H_1$ is CP odd), 
\item $\alpha_2=0$ and $\alpha_3=\pi/2$ (then $R_{13}=R_{21}=R_{22}=R_{33}=0$ and $H_2$ is CP odd),
\item $\alpha_2=0$ and $\alpha_3=0$ (then $R_{13}=R_{23}=R_{31}=R_{32}=0$ and $H_3$ is CP odd).
\eit
The corresponding regions comprise vertical green lines at the
left and right edges of the panels and green dots located in the middle
of the lower and upper sides of the panels. For our
choices of parameters the case CPC4 is never satisfied. The remaining green 
dots correspond to the case CPC5. It is worth mentioning that these green dots are not isolated points. They just appear as isolated points in our two-dimensional plots.
In the full parameter-space, these locations are parts of lower-dimensional manifolds comprising regions of CP conservation.  

\subsection{The positivity border}

As we see from Fig.~\ref{fig:singlepanel} there exist two kinds of positivity borders. One can have a non-positive/physical border and a non-positive/non-global border.
Along both kinds of borders, the potential will be flat in at least one direction, but bounded from below.
When the non-positive/physical border is crossed into the physical region, a global minimum of the potential exists, and is equal to our starting minimum (denoted ``A'').

When the non-positive/non-global border (left bottom panel in Fig.~\ref{fig:singlepanel}) is crossed into the non-global region, a global minimum of the potential exists, but our starting minimum A was not the correct one. 
Another, deeper minimum exists.

\subsection{The global minimum borders}

The region where the starting minimum $A$ is not the global one, is represented in cyan. This region can be adjacent to physical (yellow) regions and to regions where positivity is violated (pink).
The former boundaries are manifolds where spontaneous CP violation may occur. In \cite{Ivanov:2007}, it was shown that the 2HDM vacuum can be twice degenerate only when a certain symmetry 
(CP or some other symmetry) of the potential is spontaneously broken. This is consistent with our findings. We discuss these mattes in more detail below.

\subsubsection{SCPV1: $\Im\left[(m_{12}^2)^2\lambda_5^*\right]=0$}
\label{sec:scpv1}

The points satisfying SCPV1 are denoted by red curves. 
These curves separate a region where the starting minimum (A) is the global minimum (yellow) from a region where it is not. 
Thus, along the red curves, there are {\it two minima of equal depth}. 
Along the red curves our starting minimum (A) which is real exists alongside another minimum (B)
of the same depth (which is complex). The starting minimum can in the basis (\ref{Eq:v12})--(\ref{Eq:basis}) be denoted by
\begin{eqnarray}
\langle\Phi_{1}\rangle_A&=&\frac{1}{\sqrt{2}}\left( 
\begin{array}{c}
0 \\
v_1
\end{array} \right) 
\nonumber\\
\langle\Phi_{2}\rangle_A&=&\frac{1}{\sqrt{2}}\left( 
\begin{array}{c}
0 \\
v_2
\end{array} \right) 
\label{scpv1minimumA}
\end{eqnarray}
which is real. The second minimum which has the same depth is located at
\begin{eqnarray}
\langle\Phi_{1}\rangle_B&=&\frac{1}{\sqrt{2}}\left( 
\begin{array}{c}
0 \\
v_1
\end{array} \right) 
\nonumber\\
\langle\Phi_{2}\rangle_B&=&\frac{1}{\sqrt{2}}\left( 
\begin{array}{c}
0 \\
v_2e^{-i\gamma}
\end{array} \right) 
\label{scpv1minimumB}
\end{eqnarray}
where the $v_i$ are the same as for the starting minumum and $\gamma$ is the phase of $\lambda_5$, as defined by Eq.~(\ref{Eq:parameter-phases}).

In the real basis (\ref{v12realbasis}) we have:
\begin{eqnarray}
\langle\bar\Phi_{1}\rangle_A&=&\frac{1}{\sqrt{2}}\left( 
\begin{array}{c}
0 \\
v_1
\end{array} \right) 
\nonumber\\
\langle\bar\Phi_{2}\rangle_A&=&\frac{1}{\sqrt{2}}\left( 
\begin{array}{c}
0 \\
v_2e^{+i\gamma/2}
\end{array} \right) 
\nonumber
\end{eqnarray}
whereas for the other minumum we get:
\begin{eqnarray}
\langle\bar\Phi_{1}\rangle_B&=&\frac{1}{\sqrt{2}}\left( 
\begin{array}{c}
0 \\
v_1
\end{array} \right) 
\nonumber\\
\langle\bar\Phi_{2}\rangle_B&=&\frac{1}{\sqrt{2}}\left( 
\begin{array}{c}
0 \\
v_2e^{-i\gamma/2}
\end{array} \right) 
\nonumber
\end{eqnarray}
Our potential is CP invariant (as we consider the case of SCPV). Under CP
\beq
\Phi_i  \stackrel{CP}{\longleftrightarrow}  \Phi_i^*
\eeq
therefore in particular $V(\langle\Phi_{1}\rangle_A,\langle\Phi_{2}\rangle_A)=V(\langle\Phi_{1}\rangle_B,\langle\Phi_{2}\rangle_B)$. This
explains why the curve of SCPV1 separates the forbidden (non-global) and allowed (yellow) regions. 

\subsubsection{SCPV2: $\lambda_1=\lambda_2$ and $m_{11}^2=m_{22}^2$}
\label{sec:scpv2}

The red dot in Fig.~\ref{fig:singlepanel} denotes a point satisfying SCPV2.\footnote{These dots are in fact parts of a lower-dimensional manifold of the full parameter space where we have SCPV2. They appear as points only because we show a two-dimensional slice of the full parameter space.} This is also on a boundary between a forbidden and an allowed (yellow) region. 
The cyan region next to the red dot is forbidden because the starting minimum (A)
is not the global minimum. Another, deeper minimum with in general complex VEV exists there. 
In the allowed (yellow) region next to the red dot, the starting minimum (A) is the global minimum. 
A numerical study shows that for the red dot, the starting minimum (A) which is real exists alongside another minimum (B) which is also real.
These have the same depth and are related in the following way:
\begin{eqnarray}
\langle\Phi_{1}\rangle_B&=\langle\Phi_{2}\rangle_A=\frac{1}{\sqrt{2}}\left( 
\begin{array}{c}
0 \\
v_2
\end{array} \right) 
\nonumber\\
\langle\Phi_{2}\rangle_B&=\langle\Phi_{1}\rangle_A=\frac{1}{\sqrt{2}}\left( 
\begin{array}{c}
0 \\
v_1
\end{array} \right).
\label{scpv2minimum}
\end{eqnarray}

Clearly, along the border between the allowed (yellow) region and the forbidden region, on the ``back'', where there is no red curve, there are also two minima of the same depth. However, on this side, 
as opposed to the ``front'', where the red curve runs, no real basis exists, except at one single point, denoted by the red dot where CP is violated spontaneously.
The analytic expression defining the red points is given by eq.~(\ref{scpv2}). 
The values of the VEVs at the red point are specified in eq.~(\ref{scpv2minimum}).

Below, we derive analytic expresions that determine the ``back border".
As a first step, we numerically determined points along the cyan/yellow ``back border".
Then, after having located these points, the VEVs of both minima were calculated for each of the points. 
The VEV of the starting minimum (A) was of course the same value that we started out with.
The numerical evaluation of the VEV of the second minimum (B) showed that the value of $\langle\Phi_{2}\rangle_B$ 
is real along the whole ``back border". Thus, the VEVs along the border are real for both minima.
This simplifies the stationary-point equations a lot, and sets the stage for an analytical study.    

Starting with minimum A in which the vacuum is described by our input-parameters $v_1$ and $v_2$, which we here treat as known quantities, we find the following identities by using the stationary-point equations (\ref{stationary1})--(\ref{stationary4}):
\begin{eqnarray}
m_{11}^2&=&\lambda_1 v_1^2 + \lambda_{345}v_2^2 -\Re(m_{12}^2)\frac{v_2}{v_1}\nonumber\\
m_{22}^2&=&\lambda_2 v_2^2 + \lambda_{345}v_1^2 -\Re(m_{12}^2)\frac{v_1}{v_2}\nonumber\\
\Im(m_{12}^2)&=&\Im\lambda_5 v_1 v_2\label{stationaryA}
\end{eqnarray}
Here, we have used the abbreviation $\lambda_3+\lambda_4+\Re\lambda_5\equiv\lambda_{345}$. Using these identities, 
we arrive at the following expression for the value of the potential at our starting minimum A:
\begin{eqnarray}
V(\langle\Phi_{1}\rangle_A,\langle\Phi_{2}\rangle_A)&=&
-\frac{\lambda_1}{8}v_1^4-\frac{\lambda_2}{8}v_2^4-\frac{\lambda_{345}}{4}v_1^2v_2^2\label{valueatA}
\end{eqnarray}
Turning now to the second minimum (B) which the numeric study told us was real, we express it as
\begin{align}
\langle\Phi_1\rangle_B&=\frac{1}{\sqrt{2}}\left( 
\begin{array}{c}
0 \\
u_1
\end{array} \right) 
\nonumber\\
\langle\Phi_2\rangle_B&=\frac{1}{\sqrt{2}}\left( 
\begin{array}{c}
0 \\
u_2
\end{array} \right) 
\nonumber
\end{align}
where $u_1$ and $u_2$ are real, unknown quantities. The minimum B must also satisfy the stationary-point equations. Thus,
\begin{eqnarray}
m_{11}^2&=&\lambda_1 u_1^2 + \lambda_{345}u_2^2 -\Re(m_{12}^2)\frac{u_2}{u_1}\nonumber\\
m_{22}^2&=&\lambda_2 u_2^2 + \lambda_{345}u_1^2 -\Re(m_{12}^2)\frac{u_1}{u_2}\nonumber\\
\Im(m_{12}^2)&=&\Im\lambda_5 u_1 u_2\label{stationaryB},
\end{eqnarray}
and the value of the potential at minimum B becomes
\begin{eqnarray}
V(\langle\Phi_{1}\rangle_B,\langle\Phi_{2}\rangle_B)&=&
-\frac{\lambda_1}{8}u_1^4-\frac{\lambda_2}{8}u_2^4-\frac{\lambda_{345}}{4}u_1^2u_2^2.\label{valueatB}
\end{eqnarray}
By combining (\ref{stationaryA}) with (\ref{stationaryB}) and putting 
$V(\langle\Phi_{1}\rangle_A,\langle\Phi_{2}\rangle_A)=V(\langle\Phi_{1}\rangle_B,\langle\Phi_{2}\rangle_B)$,
we arrive at the following set of equations:
\begin{eqnarray}
\lambda_1 v_1^2 + \lambda_{345}v_2^2 -\Re(m_{12}^2)\frac{v_2}{v_1}
&=&\lambda_1 u_1^2 + \lambda_{345}u_2^2 -\Re(m_{12}^2)\frac{u_2}{u_1}\label{eq1}\\
\lambda_2 v_2^2 + \lambda_{345}v_1^2 -\Re(m_{12}^2)\frac{v_1}{v_2}
&=&\lambda_2 u_2^2 + \lambda_{345}u_1^2 -\Re(m_{12}^2)\frac{u_1}{u_2}\label{eq2}\\
v_1 v_2&=&u_1 u_2\label{eq3}\\
\frac{\lambda_1}{8}v_1^4+\frac{\lambda_2}{8}v_2^4+\frac{\lambda_{345}}{4}v_1^2v_2^2
&=&\frac{\lambda_1}{8}u_1^4+\frac{\lambda_2}{8}u_2^4+\frac{\lambda_{345}}{4}u_1^2u_2^2\label{eq4}
\end{eqnarray}
This is a set of four equations with only two unknown ($u_1$ and $u_2$). Combining (\ref{eq3}) and (\ref{eq4}) we solve for $u_1$ and $u_2$ (picking the only
real, positive solution {\it not corresponding to minimum A}) to get
\begin{eqnarray}
u_1=\sqrt[4]{\frac{\lambda_2}{\lambda_1}}v_2,\quad u_2=\sqrt[4]{\frac{\lambda_1}{\lambda_2}}v_1.
\end{eqnarray}
Thus,we have found that the VEVs of the second minimum (B) along the ``back border" are given by
\begin{align}
\langle\Phi_1\rangle_B&=\frac{1}{\sqrt{2}}\left( 
\begin{array}{c}
0 \\
\sqrt[4]{\frac{\lambda_2}{\lambda_1}}v_2
\end{array} \right),
\nonumber\\
\langle\Phi_2\rangle_B&=\frac{1}{\sqrt{2}}\left( 
\begin{array}{c}
0 \\
\sqrt[4]{\frac{\lambda_1}{\lambda_2}}v_1
\end{array} \right).
\nonumber
\end{align}
We see that this simplifies to the VEVs we found for minimum B in the case of SCPV2, see eq.~(\ref{scpv2minimum}).
Inserting these VEVs into either (\ref{eq1}) or (\ref{eq2}), we arrive at the same equation:
\begin{eqnarray}
\Re(m_{12}^2)=(\lambda_{345}-\sqrt{\lambda_1}\sqrt{\lambda_2})v_1v_2\label{bbordereq1}.
\end{eqnarray}
This turns out to be the equation defining the curve that constitutes the ``back border".
However, this curve can be expressed in many different ways by using the equations in (\ref{stationaryA}) to rewrite it.
After some algebra, we find that (\ref{bbordereq1}) implies
\begin{eqnarray}
\sqrt{\lambda_1}m_{22}^2-\sqrt{\lambda_2}m_{11}^2=0\label{bbordereq2}.
\end{eqnarray}
This expression does not explicitly contain $v_1$ or $v_2$, and
clearly shows that whenever we have SCPV2 ($\lambda_1=\lambda_2, m_{11}^2=m_{22}^2$), this equation is satisfied by default.

\begin{figure}[htb]
\centerline{
\includegraphics[width=16.0cm,angle=0]{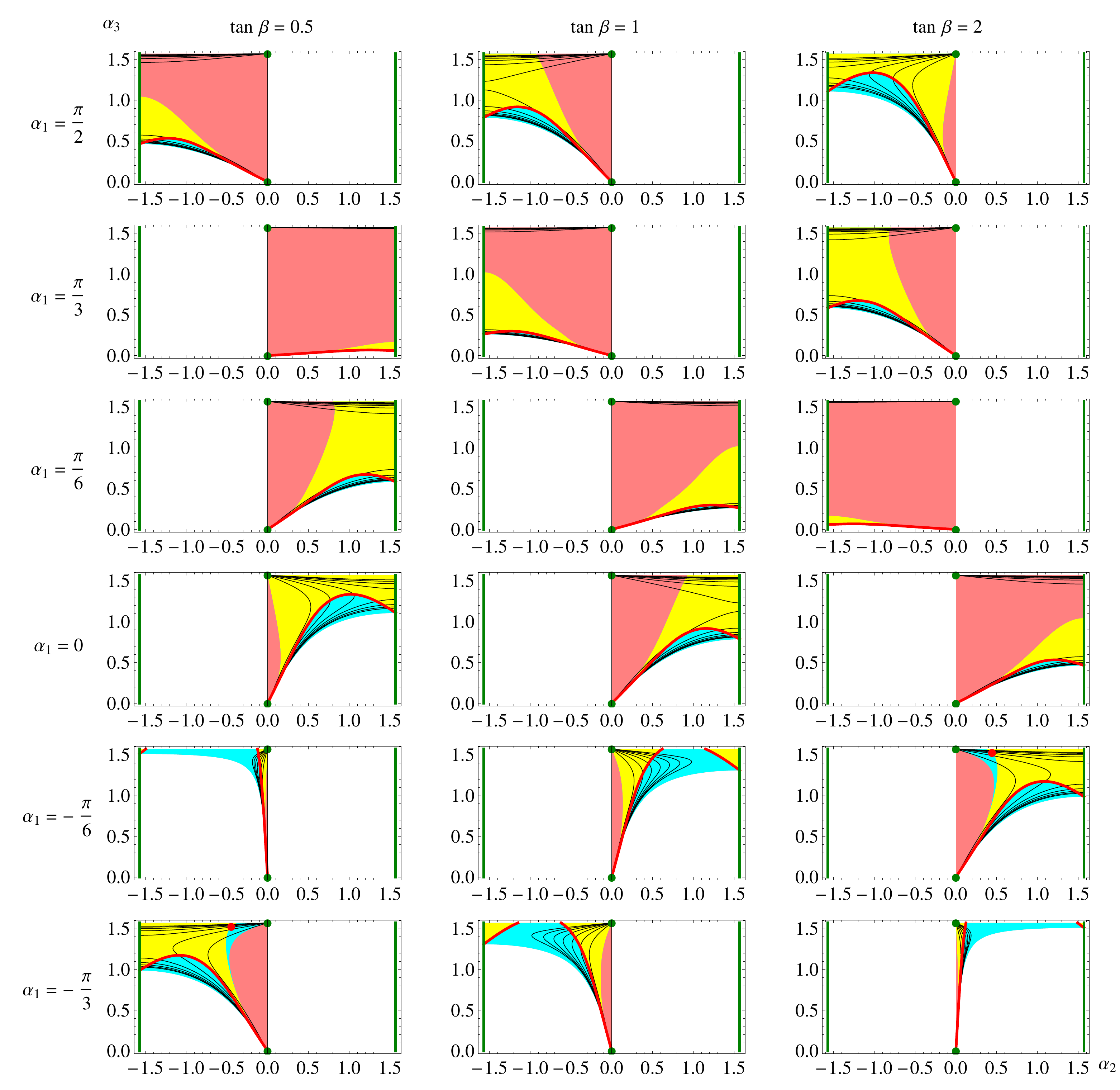}}
\caption{Similar to the bottom panels of Fig.~\ref{fig:singlepanel}, for
$M_1=125\gev$, $M_2= 200\gev$, $\mch=350\gev$, $\mu=250\gev$, 
three values of $\tan\beta$ (left to right: $0.5$, $1$, $2$) and 
six values of $\alpha_1$ (top to bottom: $\pi/2$, $\pi/3$, $\pi/6$, $0$, $-\pi/6$, $-\pi/3$).
White: Excluded because $M_3^2 < M_2^2$;
Pink: Excluded by non-positivity; Cyan: Excluded by the global minimum constraint.  
The solid black contours indicate constant values of $M_3 = 300, 400, \cdots\gev$, the curves are moving outwards from 
the vertical line $\alpha_2=0$ as $M_3$ increases.}
\label{fig:panels-rainbow1}
\end{figure}

\begin{figure}[htb]
\centerline{
\includegraphics[width=16.0cm,angle=0]{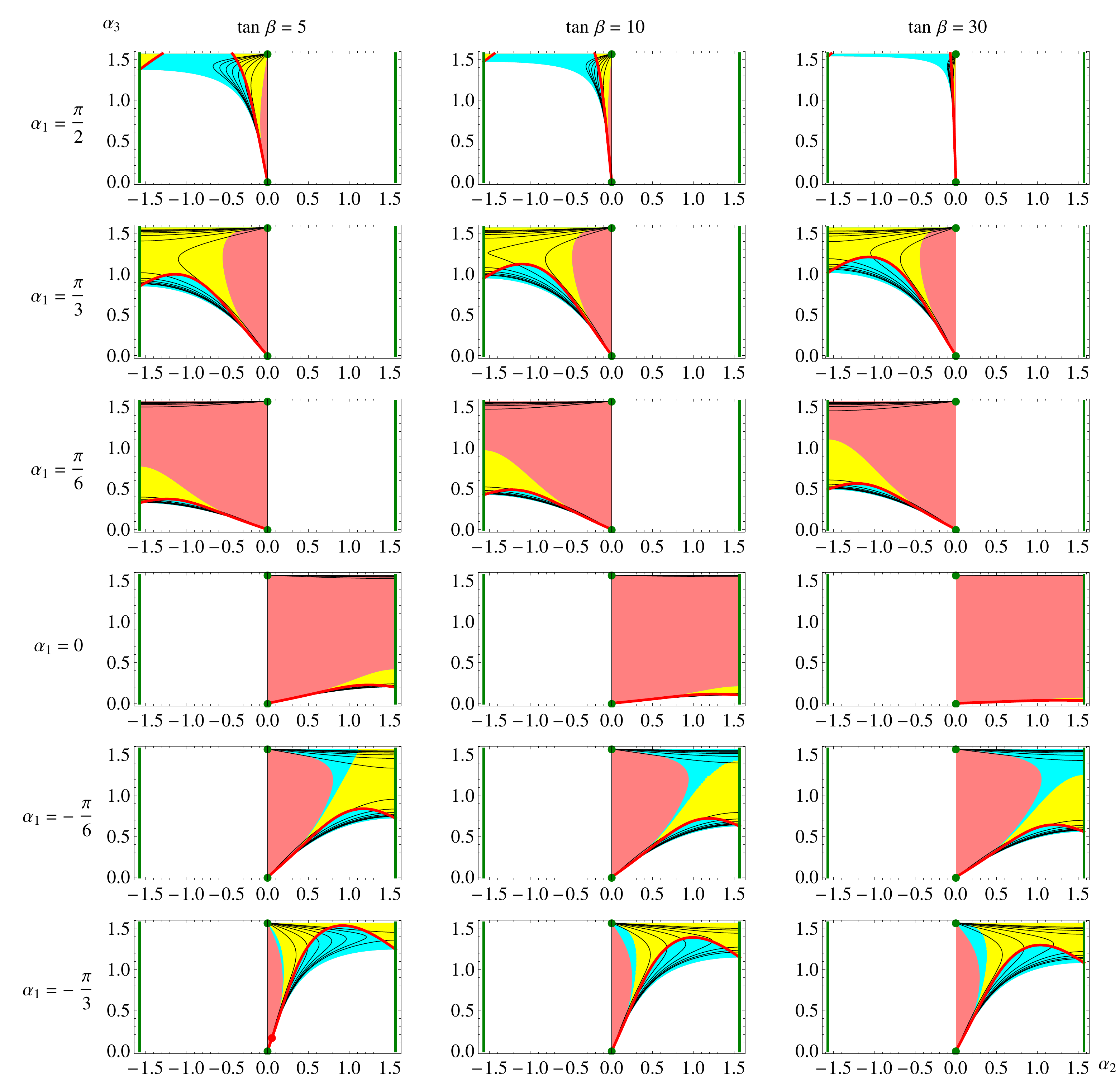}}
\caption{Similar to figure~\ref{fig:panels-rainbow1} for $\tanb=5$, 10 and 30.}
\label{fig:panels-rainbow2}
\end{figure}

\begin{figure}[htb]
\centerline{
\includegraphics[width=16.0cm,angle=0]{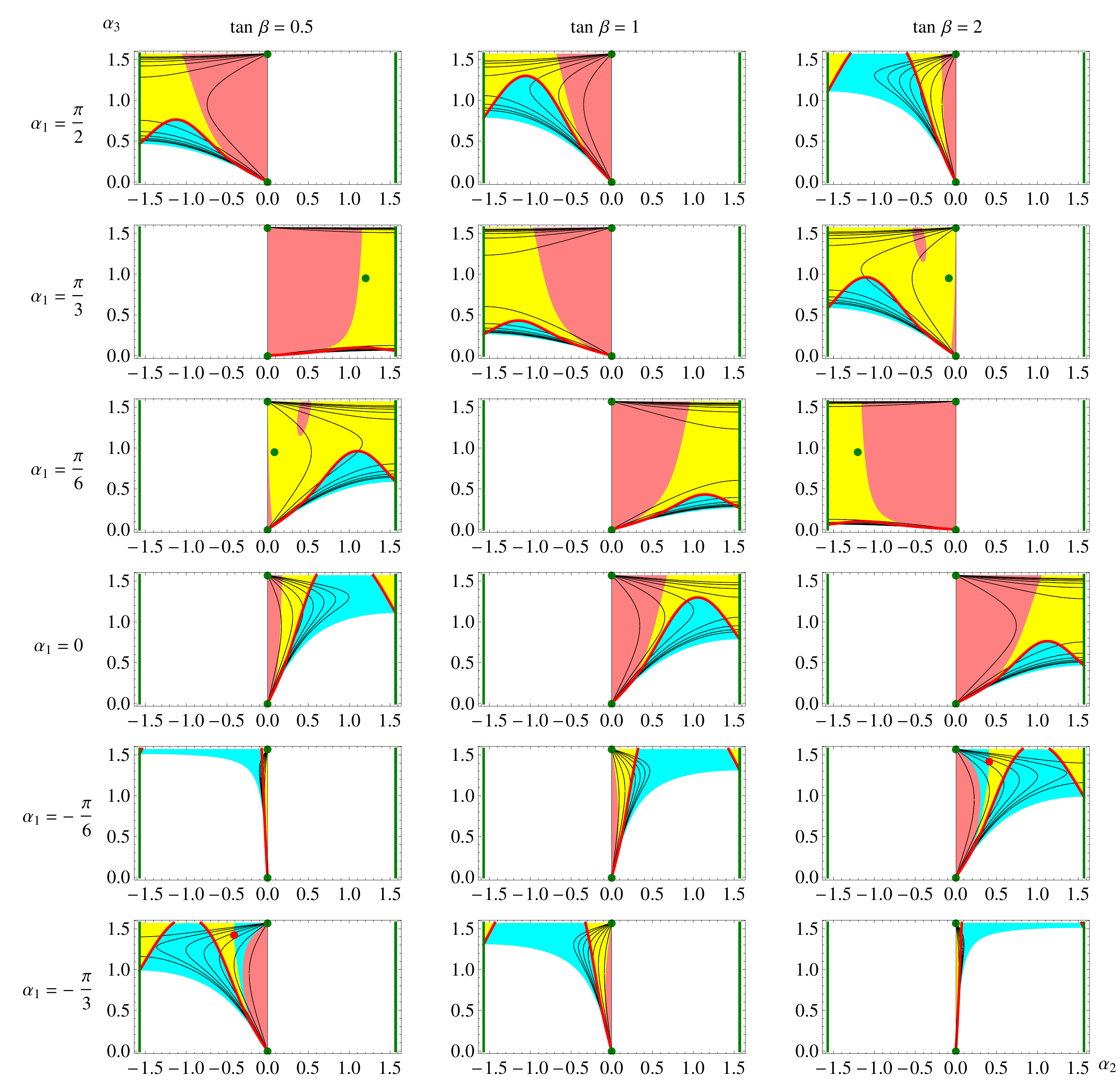}}
\caption{Similar to the bottom panels of Fig.~\ref{fig:singlepanel}, for 
$M_1=125\gev$, $M_2= 300\gev$, $\mch=500\gev$, $\mu=300\gev$, 
three values of $\tan\beta$ (left to right: $0.5$, $1$, $2$) 
and six values of $\alpha_1$ (top to bottom: $\pi/2$, $\pi/3$, $\pi/6$, $0$, $-\pi/6$, $-\pi/3$).
White: Excluded because $M_3^2 < M_2^2$;
Pink: Excluded by non-positivity; Cyan: Excluded by the global minimum constraint.  
The solid black contours indicate constant values of $M_3 = 400, 500, \cdots\gev$, the curves are moving outwards from 
the vertical line $\alpha_2=0$ as $M_3$ increases.}
\label{fig:panels-rainbow3}
\end{figure}

\begin{figure}[htb]
\centerline{
\includegraphics[width=16.0cm,angle=0]{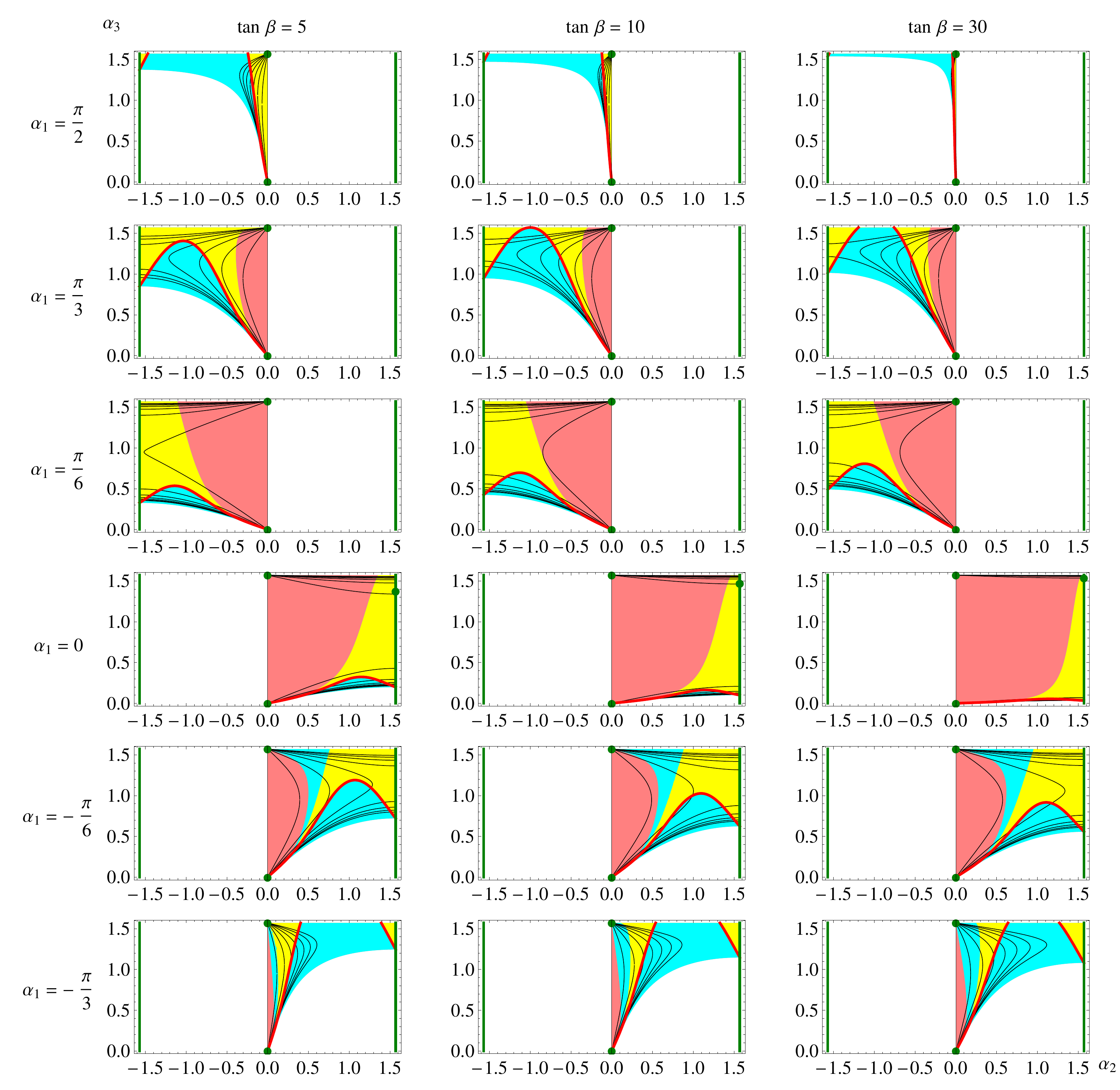}}
\caption{Similar to figure~\ref{fig:panels-rainbow3} for $\tanb=5$, 10 and 30.}
\label{fig:panels-rainbow4}
\end{figure}

We note that whenever (\ref{bbordereq2}) is satisfied, the potential is invariant under the following transformation:
\begin{eqnarray}
\Phi_1\rightarrow \sqrt[4]{\frac{\lambda_2}{\lambda_1}}\Phi_2^*,\nonumber\\
\Phi_2\rightarrow \sqrt[4]{\frac{\lambda_1}{\lambda_2}}\Phi_1^*.
\end{eqnarray}
Thus, we have identified an additional discrete symmetry of the potential along the ``back border"
that explains why we have two minima of equal depth along the curve
defined by (\ref{bbordereq2}). In fact, we easily see that the two minima A and B transform into each other under 
this transformation.

\subsection{Further illustrations}

In Figs.~\ref{fig:panels-rainbow1}-\ref{fig:panels-rainbow4} we illustrate regions
of ECPV and SCPV for the parameter choices specified at the beginning 
of Sec.~\ref{sec:case study}.

The following symmetry (discussed in section 3.1 of \cite{El Kaffas:2006nt}) can be observed in Fig.~\ref{fig:panels-rainbow1} and Fig.~\ref{fig:panels-rainbow3}:
\begin{equation}
\tan\beta\leftrightarrow\cot\beta, \quad
\alpha_1\leftrightarrow\half\pi-\alpha_1,\quad
\alpha_2\leftrightarrow-\alpha_2.
\end{equation}

The cases shown in Fig.~\ref{fig:singlepanel} correspond to the third and fifth row in the right-hand column
of Fig.~\ref{fig:panels-rainbow3}.

In all panels of Figs.~\ref{fig:panels-rainbow1}--\ref{fig:panels-rainbow4} 
positivity and the global-minimum constraint are imposed. 
We note that only one of the cases studied in Fig.~\ref{fig:singlepanel} exhibits SCPV2.
The red dot (SCPV2) appears at values of $\alpha_1$ between $-0.83$ and $-0.47$.
The red dot then appears around $(0, \pi/2)$, moves somewhat down and
to the right and then up again to disappear at $(0.49 , \pi/2)$ as $\alpha_1$ varies in this interval.

In some of these panels (for low $\tan\beta$), we note the appearance of a green dot, indicating CP conservation {\it inside} a yellow region
(recall that the yellow region denotes explicit CP violation). As already mentioned, the dot corresponds to the case CPC5.

In our examples, the high-$\tan\beta$ cases (figures~\ref{fig:panels-rainbow2} and \ref{fig:panels-rainbow4}) do not exhibit any of the isolated points (SCPV2) of spontaneous CP violation, only SCPV1 (red border between the blue and yellow regions) and explicit CP violation (yellow). Furthermore, there is no isolated point of CP conservation inside the yellow regions either. In this sense, the low-$\tan\beta$ cases have more structure.

In order to focus on CP violation, we have not exhibited the impact of other constraints. When these are imposed, significant parts of the remaining (yellow) parameter space are excluded. For the case of Type~II Yukawa couplings, see for example \cite{Basso:2012st,Basso:2013wna}.

\section{Disentangling spontaneous and explicit CP-violation}
\setcounter{equation}{0}
\label{sec:disentangling}

The ultimate goal of this study should be to propose a phenomenological
strategy that allows one to disentangle spontaneous from explicit CP-violation.
There are several comments in order, regarding that goal.

\subsection{Invariants and observables}

Any physical observable quantity must be independent of our choice of basis. This is the motivation behind giving a 
basis-independent formulation of the 2HDM. When we study a particular model and choose a particular basis suitable for the study, 
this amounts to assigning values to certain parameters, or constraining them by assuming relations between the parameters of the model. 

When we write out the full algebraic expression for an invariant quantity in the completely general 2HDM, 
without choosing any particular basis, we get an expression that is itself 
manifestly invariant. By this, we mean that applying the transformation rules for each parameter in the expression under a change of basis, we get exactly the same algebraic expression in terms of the transformed parameters.

When we write out the algebraic expressions for invariant quantities in 
a 2HDM where we have chosen a particular basis, the resulting algebraic expressions are not always manifestly invariant anymore.
So if we now apply the transformation rules for each parameter in the 
expression under a change of basis, we may get a different algebraic expression in terms of the transformed parameters.

When we perform a measurement, we determine a quantity that is basis independent.
However, in a 2HDM where a particular basis has been chosen, 
this measurement will correspond to a basis-specific algebraic expression (that is not necessarily invariant) for the measured invariant.
In this sense we may say that we {\it interpret} the measured invariant quantity as corresponding to the non-invariant 
basis-specific algebraic expression in our model.\footnote{There is a simple analogy to this in special relativity. We may measure both the energy and 
three-momentum of a particle in the rest frame of an observer even if neither energy nor three-momentum is Lorentz invariant. This is because there exist
Lorentz-invariant quantities that in the rest frame of the observer simplify to either the energy or the three-momentum of the particle.}

In our model, even without specifying the Yukawa sector, experiments will let us measure certain combinations of parameters. By combining measurements, we may thus determine 
parameters of our model. However, a parameter can only be determined from experiments if there exists an invariant (or function of invariants) that  in the model
simplifies to this parameter. Examples of observable parameters that can be determined uniquely in our model are $\lambda_3$ and $\lambda_4$, while there is a twofold ambiguity that prevents us from determining $\lambda_1$ and $\lambda_2$ uniquely. This twofold ambiguity is discussed in detail in Appendix B.

In that appendix, we arrive at the following list of nine independent observables: 
\bea
(v_1^2+v_2^2),(v_1^2-v_2^2)^2,(\lambda_1+\lambda_2),(\lambda_1-\lambda_2)^2,\lambda_3,\lambda_4,\Re\lambda_5, (\Im\lambda_5)^2,\mu^2,\label{observables}
\eea
meaning that we can determine all the parameters of the potential, except for those that would let us distinguish one doublet from the other. Of course, specifying the Yukawa sector would normally allow us to resolve the ambiguities.

Since the conditions for spontaneous CP violation are symmetric under an exchange of the two doublets,
the above-mentioned ambiguity does not prevent us from testing the origin of CP violation,
working exclusively within the bosonic sector of the model.

\subsection{Preliminaries}
For spontaneous CP violation one needs, first of all,  
at least one non-zero imaginary part of the $J_i$ invariants,
otherwise CP is conserved in the scalar sector. On top of that all $I_i$ invariants must
vanish, which is just the condition for CP invariance of the 
scalar potential. There exist various ways to detect CP violation originating from the scalar sector
experimentally, usually through measurements of certain CP asymmetries,
see e.g. \cite{Grzadkowski:2000xs,Grzadkowski:1999ye,Grzadkowski:1996pc,Gunion:1996vv,Grzadkowski:1994qk,Chang:1992tu,Wu:1999nc,BarShalom:1997sx,BarShalom:1995jb,Atwood:1995uc,Soni:1993jc,Basso:2012st,El Kaffas:2006nt,Khater:2003wq,Skjold:1994qn}. 
In fact, that is the easy part of the task, the one that 
is much more challenging is to find an experimental and simple
method to verify the conditions for CP symmetry of the potential
(\ref{scpv1}) and/or (\ref{scpv2}). Let's focus on the
condition (\ref{scpv1}). Since the condition
is formulated in terms of an invariant, it could be verified experimentally:
\bea
-\frac{4I_{2Y2Z}}{\Im J_1}=\left[4v^2\mu^2\Re\lambda_5-4\mu^4+v^4(\Im\lambda_5)^2\right]=0
\eea
In order to enable 
experimental verification of the condition for SCPV1,
one is tempted to express it through parameters that appear in
Feynman rules, e.g.\ mixing angles. That can be done
and the result is the following:
\begin{equation}
\left[4v^2\mu^2\Re\lambda_5-4\mu^4+v^4(\Im\lambda_5)^2\right]=
4\left[\Delta^2-\mu^2(M_1^2R_{13}^2+M_2^2R_{23}^2+M_3^2R_{33}^2)\right]=0,
\label{scpv1rew}
\end{equation}
where $\Delta$ is defined through 
\beq
\Delta_{ijk}=\frac{(M_k^2-M_j^2)R_{j3}R_{k3}}{(v_1R_{i1}-v_2R_{i2})}
\eeq
such that $\Delta\equiv v \Delta_{123}$. 

From equation~(\ref{scpv1rew}), two strategies are apparent:
\begin{itemize}
\item It is clear that if we can find ways to measure the three observables $\mu^2$, $\Re\lambda_5$ and $(\Im \lambda_5)^2$, we are able to test SCPV1. Determining these three observables will most probably require more than three measurements.
\item It is also clear that {\bf if} one could measure $M_1$, $M_2$, $\alpha_{1,2,3}$, $\tanb$ and $\mu^2$, then one would be able to test SCPV1. 
The neutral masses are of course observables, but $\alpha_{1,2,3}$ and $\tanb$ are not all observables due to the inability to distinguish the two Higgs doublets as we have already discussed. However, since $(\Im\lambda_5)^2=4\Delta^2/v^4$, we can conclude that $\Delta^2$ is an observable. Furthermore, $R_{i3}^2$ is unchanged under the transformation in eqs. (3.11) and (3.12) of \cite{El Kaffas:2006nt} (which amounts to interchanging the two Higgs doublets). Hence $R_{i3}^2$ are observables, and thus can be used in this approach to test SCPV1.
\item A third strategy would be to search for a combination of vertex couplings that equals the expression (\ref{scpv1rew}). Since the absolute value of vertex couplings are observables, this would outline a strategy for disentangling the CP nature of the model. But since the Feynman rules are non-trivial, this approach will most likely represent a considerable algebraic challenge.
\end{itemize}
Similar comments apply to the SCPV2 case (\ref{scpv2}). The difference is that
adopting a similar strategy, even more parameters are needed to decide
whether CP is broken spontaneously.

\subsection{Determining the potential}
We know that we cannot uniquely reconstruct the full potential and the VEVs from measurements in the scalar sector only. However, measurements in the non-fermionic sector (i.e., independently of Yukawa couplings) would be sufficient to determine the nature of the CP violation in the scalar sector. We may thus check the nature of a possible CP violation via a reconstruction of the parts of the potential that can be measured. This could proceed via measurements of masses and couplings. In principle, the masses could all be determined independently:
\begin{itemize}
\item
Measure the neutral masses $M_1$, $M_2$, $M_3$ (perhaps best done at a muon collider).
\item
Measure the charged-Higgs mass $M_{H^\pm}$ (single production via $WZ$ fusion, or pair production in $\gamma\gamma$ collisions).
\end{itemize}

The most natural attempt to determine the remaining parameters in a way that is independent of Yukawa couplings
(and therefore not sensitive to a Yukawa-coupling specific version of a 2HDM) is through measurements of 
branching ratios for Higgs bosons decaying into vector bosons:
\beq
{\rm BR}(H_i\to ZZ/W^+W^-) \sim g_{H_iZZ}^2 \sim g_{H_iW^+W^-}^2 \sim (v_1R_{i1}+v_2R_{i2})^2, \quad i=1,2,3 
\label{br2a}
\eeq
and
\beq
{\rm BR}(H_i \to H^+W^-) \sim g_{H_iH^+W^-}^2 \sim (vR_{i3})^2+(v_2R_{i1}-v_1R_{i2})^2, \quad i=1,2,3
\label{br2b}
\eeq
These 6 quantities are however not independent. For a given $i$,  the right-hand sides add to $v^2$. Since we consider $v^2$ known ($v=246~\text{GeV}$), three relations are thus removed. Furthermore, summing the right-hand sides of (\ref{br2b}) over $i$, we again get $v^2$. Thus, the couplings given by (\ref{br2a}) and (\ref{br2b}) provide two independent constraints. However the branching ratios depend on the total decay widths, therefore they are
sensitive to Yukawa couplings - the feature that we want to avoid. Therefore, in order to eliminate the total width $\Gamma(H_i)$ one has to consider ratios of branching ratios, so eventually one obtains 
only one useful constraint.
 
Here, a comment is in order. It is easy to check that $g_{H_iZZ}^2$, $g_{H_iW^+W^-}^2$ and $g_{H_iH^+W^-}^2$ 
are invariant under basis transformations, and thus observable. That implies that only quantities formed from
$\alpha_i$ and $\tanb$ that are invariants could be determined through measurements of branching ratios.

In order to test (\ref{scpv1rew}) we need three more relations. 
We could use the following decays (note that there are three equations):
\begin{equation}
{\rm BR}(H_i\to H^+H^-) \sim
\left|
2M_{H^\pm}^2\frac{v_1R_{i1}+v_2R_{i2}}{v^2}
-\mu^2\frac{v_1R_{i2}+v_2R_{i1}}{v_1v_2}
+M_i^2\frac{v_1^3R_{i2}+v_2^3R_{i1}}{v_1v_2v^2}
+\Delta\frac{v R_{i3}}{v_1v_2}
\right|^2
\end{equation}
or invoke the trilinear neutral-Higgs couplings.
In order to eliminate $\Gamma(H_i)$
one has to consider ratios, for instance one can normalize ${\rm BR}(H_i\to H^+H^-)$
to one of those two independent branching ratios discussed above, see (\ref{br2a})
and (\ref{br2b}).

Then, within the considered model, the potential could be constructed 
(up to an ambiguity irrelevant for the CP properties), 
and equation~(\ref{scpv1rew}) could be verified.

We have outlined a strategy to determine all the nine independent parameters of the potential. 
The strategy assumes that three neutral and one charged scalar are observed and their gauge and some
cubic couplings could be determined through measurements of appropriate branching ratios. Then, 
of course, it is possible to verify if CP is broken spontaneously. On the other hand 
it should be realized that equation~(\ref{scpv1rew})
contains only four invariants, one of which ($v$) is known, so one could hope that only three measurements
need to be determined: $\Re{\lambda_5}, (\Im{\lambda_5})^2$ and $\mu^2$. This observation triggers
the question: what is a minimal set of necessary measurements? For instance, if another scalar particle is 
discovered, be it $H_2$ or $H^+$, 
would it be possible to test (\ref{scpv1rew}) assuming an ideal situation such that
all couplings involving known scalar particles could be measured? 
It turns out that since it is hard to exclude non-linear relationships among couplings, it is highly non-trivial to find such a minimal set of observables that are necessary to test (\ref{scpv1rew}).
It also depends on identifying a selection of measurements that could realistically be
performed. In order to find a satisfactory solution, a detailed analysis of all the available couplings 
is needed and this is beyond the scope of the present study.

\subsection{An ideal observable?}
One could have hoped that it would be possible to find an ideal observable $\ocal_\text{CPC}$ such that
\beq
\ocal_\text{CPC} \propto \left[4v^2\mu^2\Re\lambda_5-4\mu^4+v^4(\Im\lambda_5)^2\right].
\eeq
Unfortunately that seems to be quite difficult. 

From the plots that we have presented it is clear that one could, at least
in principle, prove experimentally that CP is violated explicitly. For
that one needs to measure $M_1$, $M_2$, $\mch$, $\mu^2$. In addition $\alpha_i$ and $\tanb$ must be known (up to ambiguities). 
Then if the experimental point is located away (taking into account
experimental uncertainties) from the red curves and dots then one can conclude that
CP is violated explicitly. However, to prove that CP violation is spontaneous
is, in practice, impossible since one would need to prove that 
an experimentally allowed point in the parameter space is located {\it exactly}
either on a red curve or a red dot. Since measurements are always accompanied by
some errors, explicit CP violation would always be an option. Of course
if an experimental point would lie close to a red curve or a red dot then one could argue
that it is more natural to assume that indeed CP is violated spontaneously, since
that is connected with increased symmetry of the Lagrangian (the symmetry being, of course, 
CP itself). 

\section{Summary}
\label{Sec:summary}

We can summarize our findings in the following three points:
\begin{enumerate}
\item
The strategy we adopted uses $\tan\beta$, $\alpha_{1,2,3}$, $\mu^2$, $M_{1,2}^2$ and $M_{H^\pm}^2$ as input. Then $m_{11}^2$, $m_{22}^2$, $m_{12}^2$ and $\lambda_i$ are determined (also $M_3^2$ is fixed) adopting the stationarity conditions and relations between diagonal and non-diagonal scalar mass-squared matrices. We choose the input masses $M_{1,2}^2$ and positive $M_3^2$ so $v_1=v\cos\beta $ and $v_2=v\sin\beta $ is the location of a local minimum. Then we check numerically if the minimum is global, if it is not then it was denoted as cyan in the plots. We have also checked if the vacuum is stable by inspecting
positivity of the potential, regions where it is not the case were denoted by pink. 
\item
Parameters that correspond to SCPV lie on borders of regions for explicit CP violation (ECPV). For those parameters
there exist two vacua (related by a CP transformation) of the same depth. For fixed $M_1$, $M_2$, $\mch$, $\alpha_1$ and $\tanb$
the SCPV1 corresponds to a one-dimensional manifold (denoted by red curves) while SCPV2 corresponds to a point (red dot) as it is specified by two conditions. Red lines and red dots are located on borders between regions of ECPV (yellow) and regions where a deeper minimum exists (cyan).
\item
Red curves/dots could be approached infinitely close remaining in the region of explicit CP violation. Therefore
even if the potential parameters were known (always with some uncertainty) SCPV could be mimicked by ECPV. Of course, 
if parameters are such that the model is far from the red curves/dots, one can conclude that
CP is violated explicitly. Perhaps the simplest (theoretically) method to test SCPV1 would be to measure
$\Im\left[(m_{12}^2)^2\lambda_5^*\right]$, if that was non-zero, CP would be broken explicitly.
In spite of the twofold ambiguity that unavoidably accompanies measurements  
that are not sensitive to Yukawa couplings, the conditions for spontaneous CPV, SCPV1 and SCPV2 could be 
verified experimentally.
\end{enumerate} 

It should be stressed that in our analysis, no assumptions were made on the structure of Yukawa couplings.

\vspace*{10mm} {\bf Acknowledgements.}  
We are grateful to G. Branco and M. N. Rebelo for valuable discussions, and to the NORDITA Program ``Beyond the LHC'' for hospitality during the final stage of this work. 
We thank I.~Ivanov for bringing to our attention certain aspects of the geometric approach to the 2HDM and for comments concerning regions of CP conservation.
The research of P.O. has been supported by the Research Council of Norway. 
The work of B.G. is supported in part by the National Science
Centre (Poland) as a research project, decision no
DEC-2011/01/B/ST2/00438.

\appendix
\section{Appendix. Minimum conditions}
\label{Appendix A. Minimum conditions}
\renewcommand{\thesection}{A}
\setcounter{equation}{0}
We shall here define some notation related to minimizing the potential with respect to an independent set of variables. If we choose these to be $\Phi_1^\dagger$ and $\Phi_2^\dagger$, we get from Eq.~(\ref{Eq:v12})
\begin{align}
\frac{\partial V(\Phi_1,\Phi_2)}{\partial \Phi_1^\dagger}
&= -\frac12\left\{m_{11}^2\Phi_1
+ m_{12}^2 \Phi_2\right\} \nonumber \\
& + \lambda_1(\Phi_1^\dagger\Phi_1)\Phi_1
+ \lambda_3(\Phi_2^\dagger\Phi_2) \Phi_1
+ \lambda_4(\Phi_2^\dagger\Phi_1)\Phi_2
+\lambda_5(\Phi_1^\dagger\Phi_2)\Phi_2=0, 
\label{Eq:min-cond-1}\\
\frac{\partial V(\Phi_1,\Phi_2)}{\partial \Phi_2^\dagger}
&= -\frac12\left\{m_{22}^2\Phi_2
+ (m_{12}^2)^* \Phi_1\right\} \nonumber \\
& + \lambda_2(\Phi_2^\dagger\Phi_2)\Phi_2
+ \lambda_3(\Phi_1^\dagger\Phi_1) \Phi_2
+ \lambda_4(\Phi_1^\dagger\Phi_2)\Phi_1
+(\lambda_5)^*(\Phi_2^\dagger\Phi_1)\Phi_1=0.
\label{Eq:min-cond-2}
\end{align}
In the ``bar'red'' basis (\ref{v12realbasis}), these equations would in general have additional terms involving
$\bar\lambda_{6,7}$. 
In all, we have four conditions, two real parts and two imaginary parts must all vanish.

The real parts of these equations can be used to solve for $m_{11}^2$ and $m_{22}^2$ in terms of the $\lambda$s and the VEVs. Because of hermiticity, the imaginary parts give {\it just one} condition
\begin{equation} \label{Eq:lambda5-m12_sq}
\Im m_{12}^2=v_1v_2\Im\lambda_5.
\end{equation}

\subsection{Stationary-point equations for complex vacuum}
In the case where we have a charge-conserving minimum of the form
\begin{align}
\langle\Phi_1\rangle&=\frac{1}{\sqrt{2}}\left( 
\begin{array}{c}
0 \\
v_1
\end{array} \right) 
\nonumber\\
\langle\Phi_2\rangle&=\frac{1}{\sqrt{2}}\left( 
\begin{array}{c}
0 \\
v_2e^{i\theta}
\end{array} \right) 
\nonumber
\end{align}
the stationary-point equations are:
\begin{gather}
\lambda_1 v_1^3 + (\lambda_3 + \lambda_4 + \Re\lambda_5\cos 2\theta -\Im\lambda_5\sin 2 \theta)v_1 v_2^2 \nonumber\\
-m_{11}^2 v_1-(\Re(m_{12}^2)\cos\theta-\Im(m_{12}^2)\sin\theta )v_2=0
\label{stationary1}
\end{gather}
\begin{gather}
\lambda_2\cos\theta v_2^3 + [(\lambda_3 + \lambda_4 + \Re\lambda_5)\cos\theta- \Im(\lambda_5)\sin\theta]v_1^2 v_2 \nonumber\\
-\Re(m_{12}^2) v_1-m_{22}^2 \cos\theta v_2=0
\label{stationary2}
\end{gather}
\begin{gather}
(\Im\lambda_5\cos 2\theta + \Re\lambda_5\sin 2\theta ) v_1 v_2^2-(\Im(m_{12}^2)\cos\theta +\Re(m_{12}^2)\sin\theta)v_2=0
\label{stationary3}
\end{gather}
\begin{gather}
\lambda_2\sin\theta v_2^3+[(\lambda_3 + \lambda_4 - \Re\lambda_5)\sin\theta- \Im(\lambda_5)\cos\theta]v_1^2 v_2\nonumber\\
+\Im(m_{12}^2)v_1-m_{22}^2\sin\theta v_2=0
\label{stationary4}
\end{gather}
We note that these are necessary, but not sufficient, conditions for having a minimum of the potential. It is also worth noticing
that for $\theta=0$ equations (\ref{stationary3}) and (\ref{stationary4}) coincide.

It is also instructive to write the stationary-point conditions as two complex equations:
\begin{eqnarray}
\lambda_1 v_1^3 + \left[\lambda_3 + \lambda_4 + |\lambda_5|e^{i(\gamma+2\theta)}\right]v_1 v_2^2
-m_{11}^2 v_1-|m_{12}^2|e^{i(\alpha+\theta)}v_2&=&0
\label{Eq:min-cond-cmplx-1}\\
\lambda_2 v_2^3 + \left[\lambda_3 + \lambda_4 + |\lambda_5|e^{i(\gamma+2\theta)}\right]v_1^2 v_2
-m_{22}^2 v_2 -|m_{12}^2|e^{i(\alpha+\theta)}v_1&=&0
\label{Eq:min-cond-cmplx-2}
\end{eqnarray}

\renewcommand{\thesection}{B}
\section{Appendix. Observable parameters of the potential}
\label{AppendixB}
\setcounter{equation}{0}
In this appendix we will show how different parameters of the potential (and combinations thereof) can be written in an invariant form that is independent of 
our choice of basis. If the parameters can be written in an invariant form, it means that they are observables and can be measured. Let us start by writing the potential and the VEVs in the forms \cite{Davidson:2005,Gunion:2005ja}
\bea
V=Y_{a\bar{b}}\Phi_{\bar{a}}^\dagger\Phi_b+\frac{1}{2}Z_{a\bar{b}c\bar{d}}(\Phi_{\bar{a}}^\dagger\Phi_b)(\Phi_{\bar{c}}^\dagger\Phi_d),
\eea
and
\begin{equation}
\left<\Phi_a\right>=\frac{1}{\sqrt{2}}\left(
\begin{array}{c}0\\ v\hat{v}_a
\end{array}\right).
\end{equation}
Comparing this to (\ref{Eq:v12}) and (\ref{Eq:basis}), we find that
\bea
\hat{v}_1=\frac{v_1}{v},\quad \hat{v}_2=\frac{v_2}{v},
\eea
\bea
Y_{11}=-\frac{m_{11}^2}{2},\quad Y_{12}=-\frac{m_{12}^2}{2},\quad  Y_{21}=-\frac{(m_{12}^2)^*}{2},\quad Y_{22}=-\frac{m_{22}^2}{2}
\eea
and
\bea
&&Z_{1111}=\lambda_1,\quad Z_{2222}=\lambda_2,\nonumber\\
&&Z_{1122}=Z_{2211}=\lambda_3,\nonumber\\ 
&&Z_{1221}=Z_{2112}=\lambda_4,\nonumber\\
&&Z_{1212}=\lambda_5,\quad Z_{2121}=(\lambda_5)^*.
\eea
All other $Z_{a\bar{b}c\bar{d}}$ vanish.
In \cite{Davidson:2005,Gunion:2005ja} it is shown how to construct basis-invariant quantities from contractions between tensor indices of $V_{a\bar{b}}$, $Y_{a\bar{b}}$ and $Z_{a\bar{b}c\bar{d}}$ following a certain pattern. Using the same pattern, we are able to write parameters of our potential in an invariant way. Every quantity where we construct a scalar by contracting barred against unbarred indices in the $V$-, $Y$- and $Z$-tensors will be a basis-invariant. Let us first define the following matrices:
\bea
&&V_{a\bar{b}}=\hat{v}_a\hat{v}_{\bar{b}}^*=\frac{1}{v^2}
\begin{pmatrix}
v_1^2 & v_1v_2 \\
v_1v_2 & v_2^2
\end{pmatrix},\nonumber\\
&&Z_{a\bar{b}}^{(1)}=Z_{a\bar{c}c\bar{b}}=
\begin{pmatrix}
\lambda_1+\lambda_4 & 0 \\
0 & \lambda_2+\lambda_4
\end{pmatrix},\quad
Z_{a\bar{b}}^{(2)}=Z_{a\bar{b}c\bar{c}}=
\begin{pmatrix}
\lambda_1+\lambda_3 & 0 \\
0 & \lambda_2+\lambda_3
\end{pmatrix}.\\
&&Z_{c\bar{d}}^{(21)}=Z_{a\bar{b}}^{(2)}Z_{b\bar{a}c\bar{d}},\quad
Z_{c\bar{d}}^{(V)}=V_{a\bar{b}}Z_{b\bar{a}c\bar{d}}
\eea
Consider the invariant expressions
\bea
\frac{1}{2}\left[{\rm Tr} Z^{(2)}-\frac{{\rm Tr} \left(Z^{(2)}\right)^2-2{\rm Tr}(VZ^{(21)})}{{\rm Tr} Z^{(2)}-2{\rm Tr}(VZ^{(2)})}\right]
&=&\lambda_3\\
\frac{1}{2}\left[{\rm Tr} Z^{(1)}-\frac{{\rm Tr} \left(Z^{(2)}\right)^2-2{\rm Tr}(VZ^{(21)})}{{\rm Tr} Z^{(2)}-2{\rm Tr}(VZ^{(2)})}\right]
&=&\lambda_4.
\eea
Since these clearly invariant expressions simplify to $\lambda_3$ and $\lambda_4$ in our model, $\lambda_3$ and $\lambda_4$ are observables in our model.

The parameters $\lambda_1$ and $\lambda_2$, however, are not observables. This is due to the fact that  the labeling of the two doublets $\Phi_1$ and $\Phi_2$ is arbitrary, and interchanging the two doublets will just amount to renaming the parameters of the potential. This symmetry of the potential is written out explicitly in eqs.~(3.11) and (3.12) of \cite{El Kaffas:2006nt}. 
Therefore we will not be able to measure parameters that would let us distinguish one doublet from 
the other by performing measurements in the scalar sector only. Hence, parameters like $\lambda_1$, $\lambda_2$ and $\tan\beta$ cannot be
determined uniquely, unless one specifies the Yukawa couplings. In other words, certain combinations of parameters that are symmetric under the interchange of the two doublets are observables:
\bea
{\rm Tr}Z^{(2)}-2\lambda_3&=&\lambda_1+\lambda_2\\
2{\rm Tr}(Z^{(2)})^2-({\rm Tr}Z^{(2)})^2&=&(\lambda_1-\lambda_2)^2
\eea
Here, the fact that $\lambda_3$ has been shown to be an observable leads to the conclusion that $\lambda_1+\lambda_2$ is an observable. Together with the observable $(\lambda_1-\lambda_2)^2$ this means that one is able to determine the values of $\lambda_1$ and $\lambda_2$, but one is not able to determine which is which, i.e., there is a twofold ambiguity in the determination of these two parameters.

The same goes for $\tanb$ (or equivalently  $v_1$ and $v_2$). The quantity $v_1^2+v_2^2=v^2=(246~{\rm GeV})^2$ is invariant under a change of basis. Also consider
\bea
\frac{v^4\left({\rm Tr}Z^{(2)}-2{\rm Tr}(VZ^{(2)})\right)^2}{2{\rm Tr}(Z^{(2)})^2-({\rm Tr}Z^{(2)})^2}=(v_1^2-v_2^2)^2
\eea
The fact that $v_1^2+v_2^2$ and $(v_1^2-v_2^2)^2$ are observables (together with the fact that $v_i$ is positive) means that $v_1$ and $v_2$ can be determined up to the twofold ambiguity.

We find invariant expressions also for $\Re\lambda_5$ and $\mu^2$. Thus, these two parameters are also observables in our model. We have not substituted the invariant expressions for $(\lambda_1-\lambda_2)^2$ or $(v_1^2-v_2^2)^2$ in the following expressions. For $\Re \lambda_5$, the expression is
\bea
&&\frac{v^4}
{(\lambda_1-\lambda_2)^2\left[\right(v_1^2-v_2^2)^2-v^4]}\left[
-2{\rm Tr}(VZ^{(V)})(\lambda_1-\lambda_2)^2
-{\rm Tr}(Z^{(2)}Z^{(21)})\right.\nonumber\\
&&+3{\rm Tr}(VZ^{(21)})\left(2{\rm Tr}(VZ^{(2)})-{\rm Tr}Z^{(2)}\right)
+{\rm Tr}(VZ^{(2)})\left({\rm Tr}(Z^{(2)})^2+2{\rm Tr}Z^{(2)}{\rm Tr}Z^{(1)}\right)\nonumber\\
&&\left.
-2({\rm Tr}(VZ^{(2)}))^2({\rm Tr}Z^{(2)}+{\rm Tr}Z^{(1)})
+{\rm Tr}(Z^{(2)})^2({\rm Tr}Z^{(2)}+{\rm Tr}Z^{(1)})-({\rm Tr}Z^{(2)})^2{\rm Tr}Z^{(1)}\right]\nonumber\\
&&=\Re\lambda_5,
\eea
and for $\mu^2$
\bea
&&
\frac{v^2}
{(\lambda_1-\lambda_2)^2\left[\right(v_1^2-v_2^2)^2-v^4]}\left[
-2{\rm Tr}Y\left({\rm Tr}(VZ^{(21)})-{\rm Tr}(VZ^{(2)})
{\rm Tr}Z^{(2)}\right)\right.\nonumber\\
&&
-2{\rm Tr}(YZ^{(2)})\left(2{\rm Tr}(VZ^{(2)})-{\rm Tr}Z^{(2)}\right)
+v^2{\rm Tr}(VZ^{(V)})\left(({\rm Tr}Z^{(2)})^2-2{\rm Tr}(Z^{(2)})^2\right)\nonumber\\
&&\left.-v^2{\rm Tr}(VZ^{(2)})\left({\rm Tr}(VZ^{(21)})-{\rm Tr}(Z^{(2)})^2\right)
-2{\rm Tr}(YZ^{(21)})-v^2{\rm Tr}(Z^{(V)}Z^{(21)})\right]\nonumber\\
&&=\mu^2.
\eea
Finally, we consider $\Im\lambda_5$, which we can only determine up to a sign ambiguity because of the inability to distinguish the two doublets. Consider
\begin{equation}
\frac{4\mu^4-4v^2\mu^2\Re\lambda_5-4I_{2Y2Z}/\Im J_1}{v^4}=(\Im\lambda_5)^2.
\end{equation}
Since we have already shown that $\Re\lambda_5$ and $\mu^2$ are observables, it follows that $(\Im\lambda_5)^2$ is an observable.

In summary then, all parameters of the potential and the VEVs can in principle be measured without specifying the Yukawa sector, up to the ambiguities: (i) $\lambda_1\leftrightarrow \lambda_2$, (ii) $\Im\lambda_5\leftrightarrow-\Im\lambda_5$ and (iii) $v_1\leftrightarrow v_2$. These ambiguities are not independent. If one of them is resolved (meaning that we have been able to distinguish between the two doublets), the two others will resolve simultaneously.


\end {document}